\documentclass{aa}
\usepackage{graphicx}
\usepackage{txfonts}

\begin{document}  

\title{A Rotational and Variability Study of a Large Sample of PMS Stars in \object{NGC\,2264}}

\author{
 M. H. Lamm \inst{1}
 \and C. A. L. Bailer-Jones \inst{1}
 \and R. Mundt \inst{1}
 \and W. Herbst \inst{2}
 \and A. Scholz \inst{3}
}

\institute{Max-Planck-Institut f\"ur Astronomie, K\"onigstuhl 17, 69117 Heidelberg, Germany
 \and Van Vleck Obs., Wesleyan Univ., Middletown, CT 06459, USA
 \and Th\"uringer Landessternwarte Tautenburg, Sternwarte 5, 07778 Tautenburg, Germany} 

\offprints{M. Lamm, \email{lamm@mpia.de}}

\date{Accepted. Several figures omitted. Full version available from www.mpia-hd.mpg.de/homes/calj/ngc2264\_p1.html}


\abstract{ We present the results of an extensive search for periodic and irregular variable 
pre-main sequence (PMS) stars in the young (2--4 Myr) open cluster \object{NGC\,2264}, based 
on photometric monitoring using the Wide Field Imager (WFI) on the 2.2\,m telescope on La 
Silla (Chile). In total, about 10600 stars with $I_{\mathrm{C}}$ magnitudes between 9.8\,mag 
and 21\,mag have been monitored in our $34\arcmin \times 33\arcmin$ field. Time series data 
were obtained in the $I_{\mathrm{C}}$ band in 44 nights between Dec. 2000 and March 2001; 
altogether we obtained 88 data points per star. Using two different time series analysis 
techniques (Scargle periodogram and CLEAN) we found 543 periodic variable stars with 
periods between 0.2 days and 15 days. Also, 484 irregular variable stars were identified 
using a $\chi^2$-test. In addition we have carried out nearly simultaneous observations in 
$V$, $R_{\mathrm{C}}$ and a narrow-band H$\alpha$ filter. The photometric data enable us to 
reject background and foreground stars from our sample of variable stars according to their 
location in the $I_{\mathrm{C}}$ vs $(R_{\mathrm{C}}-I_{\mathrm{C}})$ colour-magnitude and 
$(R_{\mathrm{C}}-H\alpha)$ vs $(R_{\mathrm{C}}-I_{\mathrm{C}})$ colour-colour diagrams. We 
identified 405 periodic variable and 184 irregular variable PMS stars as cluster members 
using these two different tests. In addition 35 PMS stars for which no significant variability 
were detected could be identified as members using an $H\alpha$ emission index criterion. 
This yields a total of 624 PMS stars in NGC\,2264, of which only 182 were previously known. 
Most of the newly found PMS stars are fainter than $I_{\mathrm{C}} \simeq 15\,\mathrm{mag}$
and of late spectral type ($\ga$M2). We find that the periodic variables, as a group, have 
a smaller degree of variability and smaller $H\alpha$ index than the irregular variables. 
This suggests that the sample of periodic variables is biased towards weak-line T~Tauri
stars (WTTSs) while most of the irregular variables are probably classical T Tauri 
stars (CTTSs). We have quantified this bias and estimated that the expected fraction of WTTSs 
among PMS stars in the cluster is $77\%$. This is relatively close to the fraction of WTTSs 
among the periodic variables which is $85\%$. We also estimated the total fraction of 
variables in the cluster using only two well selected concentrations of PMS stars called 
NGC\,2264\,N \& S in which we can easily estimate the total number of PMS stars. We find that 
at least $74\%$ of the PMS stars in the cluster with $I_{\mathrm{C}}\leq 18.0\,\mathrm{mag}$ 
were found to be variable (either periodic or irregular) by our study. This number shows 
that our search for PMS stars in NGC\,2264 through extensive and accurate photometric 
monitoring is very efficient in detecting most PMS stars down to at least 
$I_{\mathrm{C}}=18.0\,\mathrm{mag}$.

}

\titlerunning {A Rotational \& Variability Study of PMS Stars in NGC\,2264}
\maketitle

\section{Introduction}
The angular momentum problem is one of the longest standing conundrums in astronomy. Briefly stated,
molecular cloud cores have four to six orders of magnitude too much angular momentum to be incorporated 
into a star. Without shedding their excess angular momentum, stars could not form. When and how they
rid themselves of angular momentum is still an open question. It is intimately connected to the issue
of disk formation and evolution, to the formation of binary systems and, during the pre-main sequence
(PMS) phase, to the magnetic interaction between stars and disks which likely regulates accretion and
drives outflows and jets (e. g. Bodenheimer, \cite{bodenh}, Bouvier et al. \cite{bouvier1997}, Stassun
et al. \cite{stassun1999}, Mathieu \cite{bob}).

Observationally, our best hope of constraining the angular momentum evolution of PMS stars is to obtain 
stellar rotation rates for objects at a variety of masses and ages. Fortunately, it is possible to do 
this photometrically because PMS stars have strong surface magnetic fields which produce large cool 
spots (e.~g. Feigelson \& Montmerle \cite{feigelson}). It is often the case that the spot pattern is 
sufficiently asymmetric and stable that the rotation period can be found by photometric monitoring (e.~g. 
Rydgren \& Vrba \cite{rydgren1983}, Herbst et al. \cite{herbst1994}, Stassun et al. \cite{stassun1999}). 
The advantage of this photometric method compared with $v\,sin\,i$ measurements is that it yields directly 
the rotation period independent of the inclination of the star and it can be very accurate even for slow 
rotating stars. Herbst et al. (\cite{herbst2001}, \cite{herbst2002}) have used this method and obtained 
rotation periods for 369 T Tauri stars (TTSs) in the $\sim\!1\,\mathrm{Myr}$ old open cluster Orion Nebular 
cluster (\object{ONC}). They found that the period distribution for higher mass stars (i.~e. $M \geq 0.25 
M_{\sun}$) is bimodal with peaks at 2 and 8 days. The bimodality is interpreted as an effect of disk-star 
interactions 
in PMS stars: Slow rotators have been magnetically locked to their disks which prevent them spinning up with 
increasing time or equivalently with decreasing radius. Magnetic coupling to an accretion disk was first 
proposed more than a decade ago as a dominant braking mechanism for PMS stars (Camenzind \cite{camenzind}; 
K\"onigl \cite{koenigl}; Shu et al. \cite{shu}). 

However, we also mention a principal limitation of the photometric method for studying the rotational periods 
among TTSs. As we will show the photometric method can much more easily measure the periods among the weak-line
T Tauri stars (WTTSs) but only for a fraction of the classical T Tauri stars (CTTSs) due to their higher 
irregular variability. This ``photometric noise'' often prevents the detection of a periodic signal in the 
photometric data. Therefore any photometric monitoring of a young open cluster (like the ONC or NGC\,2264) 
will be biased towards WTTSs in the sense that for these stars the fraction of measurable periods will 
be higher.

From the results of the rotational study in the ONC several questions naturally arise: 1) Is the period 
distribution similar in other clusters (i.~e. does environment play a role)? 2) How many PMS stars interact 
with their disks and how strongly does this affect the angular momentum evolution? 3) How does the period 
distribution evolve with time (i.~e. do most stars decouple from their disks on time scales of the age of 
an ONC star and spin up with conserved angular momentum or do disks and disk-locking persist for longer 
time scales)? To answer these questions it is necessary to measure rotation periods of large samples of 
PMS stars in clusters with different ages.
 
Aside from the ONC, the open cluster \object{NGC\,2264} is perhaps the best target for a detailed rotational study, 
since it is sufficiently nearby (760~pc, Sung et al. \cite{sung}), fairly populous, and with an estimated age of 
2 -- 4\,Myr (Park et al. \cite{park}) it is about a factor of 2 -- 4 older than the ONC. 

Prior to our study a few monitoring programs have been carried out in \object{NGC\,2264} (e.g. Kearns \& 
Herbst \cite{KH}) which  altogether yielded  only about 30 published rotation periods for PMS stars. 
Vogel \& Kuhi \cite{vogel1981} and Soderblom et al. (\cite{soderblom1999}) reported rotational 
velocities ($v\,\sin\,i$) for in total about 60 low mass NGC\,2264 stars. An early variability study 
has been carried out by Nandy \& Pratt (\cite{nandy1972}) who identified 26 optical variable stars in 
the cluster. Neri, Chavarr\'{\i}a, \& de Lara (\cite{neri1993}) reported optical and near-infrared 
photometry for 50 stars in NGC\,2264 and detected for $70\%$ of these stars optical variability. Due
to the limiting sensitivity of these studies most of the stars have spectral types earlier than K5. 

About 200 PMS stars and some 20 PMS candidates have been identified in NGC\,2264 prior to our study using 
a variety of methods including $H\alpha$ spectroscopy (Herbig \cite{herbig}; Ogura \cite{ogura}; Marcy 
\cite{marcy1980}; Rydgren \cite{rydgren1979}), $H\alpha$ narrow band photometry (Sung et al. \cite{sung}; Park 
et al. \cite{park}) or ROSAT X-ray flux measurements (e.g. Flaccomio et al. \cite{flaccomio}). A PMS membership 
catalogue that summarises (most) these results is available in Park et al. (\cite{park}). 

Vasilevskis, Sanders, \& Balz (\cite{vasil1965}) have carried out a proper motion study of 245 stars in the 
region of the cluster. However, the suggested membership probabilities must be handled with caution because several
stars with strong H$\alpha$-emission which clearly must be members have low membership probability while other 
(non-emission) stars with high membership probability are clearly no members (see e.~g. the H$\alpha$-emission study 
by Rydgren \cite{rydgren1979}).
King (\cite{king1998}) and Soderblom et al. (\cite{soderblom1999}) have in total determined lithium abundances 
of 35 cluster members. Recently, circumstellar disk candidates have been identified in NGC\,2264 by Rebull et al. 
(\cite{rebull2002}). 

In this paper we present the results of an extensive photometric monitoring program in NGC\,2264, in which we 
monitored $\sim 10600$ stars over a very broad magnitude range ($9.8 \leq I_{\mathrm{C}}
\leq 21$). This extensive monitoring program allowed us to discover about 400 new PMS stars in the cluster and in 
addition we could measure the rotation periods of 405 PMS stars down to the substellar limit.

In the following section we describe the observational and data reduction strategy. Sect.~\ref{photometrie} describes 
the methods we have used for obtaining absolute and relative photometry. In Sect.~\ref{timeseries} we describe the 
time series analysis. In Sect.~\ref{member} we discuss in detail how we selected the PMS stars among the periodic and 
irregular variables we have found. The nature of the stellar variability is examined in Sect.~\ref{natvari}. The 
identification of additional PMS members for which no variability could be detected is described in Sect.~\ref{PMScand}.
In Sect.~\ref{comleteness} we estimate the fraction of variable PMS stars and the completeness level of our PMS sample.
In Sect.~\ref{fractions} we investigate if our sample of periodic variables is representative for all cluster members.
Our major conclusions are summarised in Sect.~\ref{summary}. The obtained extensive rotational period data and possible 
consequences for the disk-locking scenario will be discussed in a separate paper (see Lamm et al. \cite{lamm}, hereafter
Paper II).

\section{Data acquisition \& reduction}\label{da}

\begin{figure}
  \resizebox{\hsize}{!}{\includegraphics{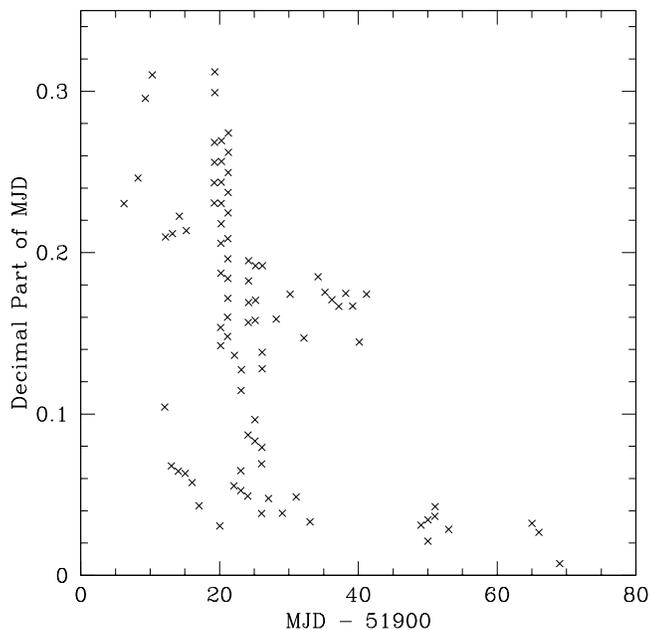}}
	\caption{Distribution of observation times of the $I_{\mathrm{C}}$ band time series (between 30 Dec. 2000 
		 and 01 Mar 2001). The time (expressed as a fraction of a day) as a function of the modified Julian
		 day (MJD) is shown.}	
	\label{obsdates}
\end{figure}

The photometric monitoring program described here was carried out in the $I_{\mathrm{C}}$ band on 44 nights 
during a period of two month between 30 Dec. 2000 and 01 Mar. 2001 with the Wide Field Imager (WFI) on the 
MPG/ESO 2.2\,m telescope on La Silla (Chile). The WFI consists of a mosaic of four by two CCDs with a total 
array size of $8\,\mathrm{k}\times 8\,\mathrm{k}$k. The field of view is $34\arcmin\,\times\,33\arcmin$ and 
the scale is 0.238\arcsec/pixel. In Fig.~\ref{bild} we show a 500\,sec $R_{\mathrm{C}}$ exposure of our 
observed field from which the positions of the eight CCDs is evident. To avoid highly saturated images and 
light scattering from the very bright star \object{S Mon} ($V=4.7\,\mathrm{mag}$) this star was located near 
the northern end of the 23\farcs 3 (96 pixels) wide central gap between two chips. The central position on 
the sky was close to $\mathrm{RA(2000)}=6^h\,40^m\,59^s$ and $\mathrm{DEC(2000)}=9\degr\,38\arcmin\,59\arcsec$ 
for most frames. The typical deviation from this nominal position is 2$\arcsec$. About $5\%$ of the imaged 
area is lost due to the gaps between the chips and the small dithering of the frames. In order to increase 
the dynamical range of the observations, three consecutive exposures of 5~sec, 50~sec and 500~sec were taken 
with the $I_{\mathrm{C}}$ filter. This series of three exposures is defined here as one data point. In total 
we obtained 110 data points and between 1 and 18 data points per night. The observing time distribution of our 
time series is shown in Fig.~\ref{obsdates}. The typical seeing (measured by the FWHM of the PSF) in these 
images was of the order of $0\farcs 8$--$1\farcs 2$. 

\begin{figure*}
  \centering
  \caption{Reproduction of a 500\,sec $R_{\mathrm{C}}$ band exposure of our observed field. The area imaged 
	   by the individual chips a -- h is quite evident. The bright star near the northern end of the gap
	   between chip b and c is S Mon (V=4.7). In the  second chip g the famous cone nebular is visible,
	   which is hardly evident on our $I_{\mathrm{C}}$ band images as is the case for the nebulosity in
	   the centre of chip e.}	
  \label{bild}
\end{figure*}

In addition to the $I_{\mathrm{C}}$ observations we observed the cluster through V and $R_{\mathrm{C}}$ 
filters on six nights during the 2000/2001 season. In $R_{\mathrm{C}}$ the exposure times were 5~sec, 
50~sec and 500~sec, while the exposure times through the V filter were set to 5~sec, 60~sec and 720~sec. 
$I_{\mathrm{C}}$ observations were obtained directly before or after an observation in V or $R_{\mathrm{C}}$. 
This allows us to determine the ($\mathrm{V}-I_{\mathrm{C}}$) and ($R_{\mathrm{C}}-I_{\mathrm{C}}$) colours 
from nearly simultaneous measurements in the two filters, i.~e. the colours should not be affected by 
variability to any substantial extent. 

In order to improve the absolute photometry, additional $V$, $R_{\mathrm{C}}$ and $I_{\mathrm{C}}$ WFI 
data were obtained between October 2001 and March 2002 using the same exposure times. In order to search 
for $H\alpha$ emission, three consecutive $H\alpha$ observations of 150~sec, 1200~sec and 1200~sec exposure 
time were taken directly before or after the observations in $R_{\mathrm{C}}$ in order to obtain nearly 
simultaneously observations in these two filters for a determination of the ($R_{\mathrm{C}}-\mathrm{H}\alpha$) 
colour. The used $H\alpha$ filter has a central wavelength $\lambda = 6588\,\mathrm{\AA}$ and FWHM = 74\,\AA.
A detailed list of observations in filters other than $I_{\mathrm{C}}$ is given in Table~\ref{observations}. 
\begin{table}
\centering
\caption{Dates of observations with the $V$, $R_{\mathrm{C}}$ and $H\alpha$ filter for the two observing seasons
	 2000-2001 and 2001-2002. Listed are the number of data points (i.~e. three images of three different
	 exposure times) for each night in a given filter. The observations in V and R$_C$ were take directly 
	 before or after a set of three images in $I_{\mathrm{C}}$.
	}
  \begin{tabular}{r c c c}
	\hline
	\hline
	\multicolumn{1}{c}{Date of} & \multicolumn{3}{c}{data points in} \\
	\multicolumn{1}{c}{observation} & $V$ & $R_{\mathrm{C}}$ & $H\alpha$ \\
	\hline
	 9 Jan 2001 & 1 & -- & --\\
	10 Jan 2001 & 1 & -- & --\\	
	11 Jan 2001 & 1 & -- & --\\	
	12 Jan 2001 & 1 & 1 & --\\	
	15 Jan 2001 & -- & 1 & --\\	
	16 Jan 2001 & -- & 1 & --\\
	\hline	
	25 Nov 2001 & 1 & 1 & 1\\	
	11 Dec 2001 & 1 & -- & --\\
	 4 Jan 2002 & 1 & 1 & 1\\	
	13 Jan 2002 & 1 & -- & --\\	
	17 Jan 2002 & 1 & -- & --\\	
	 2 Mar 2002 & 1 & 1 & 1\\	
	\hline
	total & 10 & 6 & 3\\
	\hline
  \label{observations}
  \end{tabular}
\end{table}

The image processing was done for each of the eight WFI chips separately using the standard IRAF tasks. A 
one-dimensional bias was subtracted from each frame using the overscan region in each frame. A small 
two-dimensional residual remained which was removed using zero integration time (bias) frames. For images 
through V, $R_{\mathrm{C}}$ and H$\alpha$ filters the variable sensitivity across the chips was corrected 
using the median combination of typically 10\,--\,15 twilight flats taken in the beginning and the end of 
two or three different nights. 
In $I_{\mathrm{C}}$ we used illumination-corrected dome flats for flatfielding the science images. This was 
necessary because the $I_{\mathrm{C}}$ twilight flats showed interference fringes caused by narrow band emission 
from the Earth's atmosphere. The illumination-corrected flats were created in the way described by 
Bailer-Jones \& Mundt (\cite{caljm}). Since the science images in the $I_{\mathrm{C}}$-band were affected by 
fringing too, it was essential that these fringes be removed. We note that fringing is an additive effect
and therefore the fringe pattern has to be subtracted from the science images. Since the fringe pattern 
was found to be stable over a given observation run we created only one fringe image for each chip per 
observing season (for more details see Bailer-Jones \& Mundt, \cite{caljm}).

\section{Photometry and Astrometry}\label{photometrie}

Our goal was to perform relative and absolute photometry for every star in our field with a sufficient signal-to-noise 
ratio, i.~e. $S/N=10$ or better. The relative and absolute photometry is described in the following subsections 
separately. As a common first step we had to identify the positions of all sources in our field. Therefore we first 
created a source list for all 8 chips employing the DAOFIND task, with a separate list for the 5\,sec, 50\,sec and 
500\,sec exposures. For this source search we selected for each of the three 5\,sec, 50\,sec and 500\,sec 
exposures one frame with a position very close to the nominal central location of all frames. These images we designate 
here as reference frames. The number of sources detected in each chip of these reference images is listed in 
table~\ref{nos}. 
The coordinates of each source in all other frames were calculated by applying a derived linear offset between the 
reference images and these frames. The offsets were calculated for each chip separately by measuring the pixel 
position of a bright star close to the chip's centre in all frames. 
For the whole observed field the offsets were calculated by using eight bright stars. During the aperture photometry 
all sources were re-centered (see below) and therefore the offsets calculated in that way are sufficient for identifying 
the stars in the all other frames.
\begin{table}
\centering
\caption{Number of detected sources in the various chips of the reference images.} 
  \label{nos}
  \begin{tabular*}{88mm}{@{\extracolsep\fill}r|@{\extracolsep\fill}r@{\extracolsep\fill}r@{\extracolsep\fill}r@{\extracolsep\fill}r@{\extracolsep\fill}r@{\extracolsep\fill}r@{\extracolsep\fill}r@{\extracolsep\fill}r@{\extracolsep\fill}r} 
	\hline
	\hline
	\multicolumn{1}{@{\extracolsep\fill}c|}{Exposure} & \multicolumn{8}{@{\extracolsep\fill}c}{Number of sources in Chip} &  \\
	\multicolumn{1}{@{\extracolsep\fill}c|}{time} & a & b & c & d & e & f & g & h & total\\
	\hline
	  5~sec & 505 & 268 & 445 & 311 & 405 & 278 & 286 &  334 & 2832\\
 	 50~sec & 864 & 447 & 586 & 883 & 814 & 546 & 432 & 648 & 5220\\
	500~sec & 2220 & 900 & 1187 & 1535 & 1754 & 1054 & 654 & 1645 & 10940\\ \hline
  \end{tabular*}
\end{table}

The IRAF tasks CCXYMATCH and CCMAP were used to cross-identify our sources in the reference frames with 
objects in the USNO-A2.0 catalogue (Monet et al. \cite{monet}) and to calculate the plate transformations to the sky 
positions. We typically used 70 reference stars (minimum 30, maximum 133) for the calculation of the plate solutions 
in each chip. Using these plate solutions the final J2000 coordinates of our sources were calculated with the 
task CCTRAN. These coordinates are listed in Table~\ref{ergebnis}. The RMS of the deviations between the fit and 
the coordinates of the reference stars is typically 0\farcs1.

\subsection{Relative photometry}\label{relphot}

All the variability studies described here are based on differential photometry relative to a set of non-variable 
reference stars. Before the final analysis and after various tests, we rejected some frames from the further analysis.
These were usually frames that were taken under poor transparency conditions, poor seeing or with a bright 
background due to scattered moon light. In total we rejected 22 of the 110 data points before the final analysis, so 
that relative photometry was performed on 88 remaining data points. 

The DAOPHOT/APPHOT task was used to measure the brightness of each object in the source lists. The aperture was 
chosen to be 8~pixels ($1\farcs 9$) in diameter for all measurements in order to maximise the signal-to-noise ratio. 
The sky was calculated for each source separately as the median of an annulus with an inner diameter of 30~pixel 
($7\farcs 1$) and a width of 8~pixel centred on the source. 
For the calculation of the sky value we used a sigma clipping rejection criterion with a $3\sigma$-threshold.
During the measurement the sources were re-centred. Sources with any pixel entering the $\leq 1\%$ nonlinearity 
region were rejected.

The differential magnitudes for all sources were formed as follows. First a set of non-variable comparison stars 
was selected from all sources in each chip for each of the three different exposures. Comparison stars were chosen 
according to the following criteria: 1) present on every of the 88 frames, 2) isolated from other sources and the corners 
of the field (see below), 3) the photometric error given by APPHOT (i.~e. Poisson error in the source and scatter in the 
background) is on average less than 0.01~mag. Only stars that passed all of these three tests were selected.
Possible variables among the candidate comparison stars were identified by comparing the flux of each candidate with the 
mean flux of all other candidates at the different epochs. Stars that show the largest variability were rejected from the 
candidate sample and the procedure was performed again with the remaining stars. After a few iterations typically ten 
non-variable comparison stars were identified in each chip so that the mean standard deviation in the relative light 
curves of the comparisons stars (relative to the other comparison stars) was typically $\sigma = 0.009\,\mathrm{mag}$. 
In the best cases we achieved a mean standard deviation of $\sigma = 0.006\,\mathrm{mag}$ which is therefore the 
maximum precision we can expect for the relative photometry. This limit is most likely set by flat fielding errors and 
not by photon noise. 

The selected comparison stars were used for the calculation of the relative magnitude $m_{\mathrm{rel}}(t_i)$ of each 
source in the field at the different epochs $t_i$. The relative magnitude and its error $\delta m_{\mathrm{rel}}(t_i)$ 
were calculated as described in Bailer-Jones \& Mundt (\cite{caljm}). We note that the mean was subtracted from 
each light curve so that $\sum_i m_{\mathrm{rel}}(t_i) = 0$.

In Fig.~\ref{meanerr} we show the mean error $\overline{\delta m_{\mathrm{rel}}}$ in a single 
measurement as a function of magnitude for the stars measured with the three different exposure times (for the magnitude 
determination see section~\ref{absphot}). As an example we show the light curves of four stars in Fig.~\ref{phase_lc}.
\begin{figure}
  \resizebox{\hsize}{!}{
}
	\caption{The mean error $\overline{\delta m_{\mathrm{rel}}}$ in a single measurement for the three different 
		 exposure times of 5s (top), 50s (mid) and 500s (bottom) as a function of $I_{\mathrm{C}}$ magnitude. 
		 The shaded regions indicate which exposure time we used for the further analysis of a star with a 
		 given magnitude (see text). Note the different magnitude scales in the three plots.}	
	\label{meanerr}
\end{figure}
\begin{figure*}
  \centering
	  \caption{Examples for relative light curves (mean subtracted) of four stars and the resulting phased light 
		   curve for each of the four stars. The light curves were phased with the period that we found by the 
		   Scargle periodogram analysis technique. For each star we also list the Number (no.), the mean 
		   $I_{\mathrm{C}}$ magnitude and the period P.}
  \label{phase_lc}
\end{figure*}

\subsection{Absolute photometry}\label{absphot}

\begin{table*}
\centering
\caption{Comparison of our photometry with those of other authors. We list the offsets in the sense results of other author 
	minus our results. We also show the number of stars which we used for the comparison with other studies. These
	numbers differ for a single publication because not all magnitudes or colours were available for all stars. Stars 
	with close-by neighbours were not not used for the calculation of the offsets, since these stars are not separated in 
	the other studies.
	} 
  \begin{tabular}{l c r| c r |c r}
	\hline
	\hline
	\multicolumn{1}{c}{Publication} & \multicolumn{1}{r}{No. of stars}& \multicolumn{1}{c|}{$I_{\mathrm{C}}$} & \multicolumn{1}{r}{No. of stars}& \multicolumn{1}{c|}{($R_{\mathrm{C}} - I_{\mathrm{C}}$)} & \multicolumn{1}{r}{No. of stars}& \multicolumn{1}{c}{($V - I_{\mathrm{C}}$)}\\
	\hline
	Rebull et al.(2002)    & 1344& $ 0.003\pm 0.002$ & 1047     & $-0.019\pm 0.002$ &  767 & $-0.010\pm 0.002$ \\
	Park et al. (2000)     &  147& $-0.009\pm 0.012$ & \dotfill & $\dotfill       $ &  147 & $-0.041\pm 0.002$ \\
	Flaccomio et al. (1999)&  236& $ 0.070\pm 0.003$ &  236     & $-0.072\pm 0.008$ &  221 & $-0.147\pm 0.007$ \\
	Sung et al. (1997)     &  117& $-0.000\pm 0.002$ & \dotfill & $\dotfill       $ &  116 & $ 0.001\pm 0.010$ \\
	\hline
  \label{phovgl}
  \end{tabular}
\end{table*}

In order to obtain additional constraints on the nature of the observed stars, in particular on their PMS membership, we
obtained absolute photometry in $V$, $R_{\mathrm{C}}$ and $I_{\mathrm{C}}$. As we will outline below all of our 
photometry is done relative to secondary standards in our observed field. The result of the absolute photometry is
reported in Table~\ref{ergebnis}.

Many of the stars in the field are expected (and found, see Section~\ref{timeseries}) to be variable and the peak-to-peak 
variations of the variable stars are in most cases $\la 0.2\,\mathrm{mag}$ in $I_{\mathrm{C}}$ (see Fig.\,\ref{sigma_vert}). 
Therefore a single measurement of the magnitude of a star depends on the phase and amplitude of the star's brightness 
modulation. On the other hand single measurements of the colour of a PMS star with cool (or hot) surface spots also 
differ at different epochs. That is because the spectral energy distribution of the light coming from the area of the spot 
differs from the spectral energy distribution of the light radiated from the other part of the stellar atmosphere 
because of the different effective temperatures. The difference between the two energy distributions is wavelength 
dependent and larger for shorter wavelengths. Therefore, if cool spots cause the variability of a star the peak-to-peak 
variation in its light curve increases to shorter wavelength, i.~e. it is larger in the $V$-band and smaller in the 
$I_{\mathrm{C}}$ band. Observations of T Tauri stars confirm this expected colour behaviour of the stars (e.g. Vrba 
et al. \cite{vrba}; Herbst et al., \cite{herbst1994}). The relations between the peak-to-peak variations $\Delta V$, 
$\Delta R_{\mathrm{C}}$ and $\Delta I_{\mathrm{C}}$ in the different filters are given by the colour slopes 
\begin{eqnarray*}
   S_R=\frac{d(\Delta R_{\mathrm{C}}-\Delta I_{\mathrm{C}}) }{d(\Delta I_{\mathrm{C}})}
   \hspace{1cm}  \mathrm{and}  \hspace{1cm} 
   S_V=\frac{d(\Delta V_{\mathrm{C}}-\Delta I_{\mathrm{C}}) }{d(\Delta I_{\mathrm{C}})}.
\end{eqnarray*}
Herbst et al. (\cite{herbst1994}) calculated these slopes for several WTTSs with a mean value of $S_R = 0.31 \pm 
0.23$ and $S_V = 0.55 \pm 0.36$. Since the typical peak-to-peak variations we found in our target stars are of the order 
of $\Delta I_{\mathrm{C}}\simeq 0.2$ we expect maximal colour changes of about $\Delta(R_{\mathrm{C}}-I_{\mathrm{C}})\simeq
0.06$ and $\Delta(V-I_{\mathrm{C}}) \simeq 0.1$ but in some extreme cases the variations in the colour may be higher.

To take into account these magnitude and colour changes we determined the $I_{\mathrm{C}}$ magnitude and the ($V -
I_{\mathrm{C}}$) and ($R_{\mathrm{C}}-I_{\mathrm{C}}$) colours at different epochs. The colours and magnitudes we list 
in Table~\ref{ergebnis} are average values of these different measurements.  The dates of the employed measurements are 
listed in Table~\ref{observations}. 

For the determination of the average $I_{\mathrm{C}}$ magnitude of each star we used only images which were taken during 
the first observing season in Jan. 2001. In this season we got nearly simultaneous measurements in $I_{\mathrm{C}}$ and
$R_{\mathrm{C}}$ (i.~e. three images in $I_{\mathrm{C}}$ with different exposure times followed by three images of 
different exposure times in $R_{\mathrm{C}}$) or nearly simultaneous measurements in $I_{\mathrm{C}}$ and $V$ at seven 
different epochs (see Table~\ref{observations}). 
The final averaged ($R_{\mathrm{C}}-I_{\mathrm{C}}$) colours were calculated from nearly simultaneous measurements in 
$R_{\mathrm{C}}$ and $I_{\mathrm{C}}$ at the six different epochs during both observing seasons listed Table~\ref{observations}. 
The averaged ($V-I_{\mathrm{C}}$) colours were calculated from nearly simultaneous measurements in $V$ and $I_{\mathrm{C}}$ 
at the 10 different epochs in both observing seasons listed in Table~\ref{observations}. 
The transformation of our instrumental $V_{\mathrm{instr}}$, $R_{\mathrm{C,instr}}$ and $I_{\mathrm{C,instr}}$ magnitudes into 
the true ($V-I_{\mathrm{C}}$) and ($R_{\mathrm{C}}-I_{\mathrm{C}}$) colours and $I_{\mathrm{C}}$ magnitude is outlined below.
It was done before the averaging and for each epoch, each chip and each exposure time separately using a linear transformation.

Since we did not observe any flux standards during any of the observing seasons we had to use secondary standard stars 
located in our field for calibrating our measurements. The photometric calibration coefficients were determined using 
magnitudes and colours of stars in NGC\,2264 measured by Rebull et al. (private communication). Their photometric data 
include the photometry of PMS stars in the cluster from Rebull et al. (2002) and in addition unpublished photometry of
foreground and background objects in their observed field. This extended dataset including the unpublished data, compared 
to a dataset consisting only of PMS stars, has the advantage of a smaller fraction of variable stars. Our objects were 
cross-identified with the 2924 objects in the extended Rebull et al. catalogue that are located in our field. The 2227 
($76\%$) stars which we could identify we used as secondary photometric standard stars. 

We transformed our instrumental magnitudes $I_{\mathrm{C,instr}}$ into the Cousins $I$ system by applying the linear 
transformation equations of the form 
\begin{eqnarray*}
    I_{\mathrm{C}} = a_1 + b_1 \times (R_{\mathrm{C,instr}}-I_{\mathrm{C,instr}}) + I_{\mathrm{C,instr}} \hspace{1cm} 
    \mathrm{and}
\end{eqnarray*}
\begin{eqnarray*}
    I_{\mathrm{C}} = a_2 + b_2 \times (V_{\mathrm{instr}}-I_{\mathrm{C,instr}}) + I_{\mathrm{C,instr}}
\end{eqnarray*}
to the measurements in $I_{\mathrm{C}}$ which were nearly simultaneous with one of the other two filters $V$ or 
$R_{\mathrm{C}}$. The coefficients $a_1$ and $b_1$ were calculated simultaneously using the secondary photometric 
standard stars by applying a linear least squares fit to the points in the ($I_{\mathrm{C,Rebull}} - 
I_{\mathrm{C,instr}}$) vs ($R_{\mathrm{C,instr}}-I_{\mathrm{C,instr}}$) plane, where $I_{\mathrm{C,Rebull}}$ is 
the magnitude in our secondary standard catalogue. The coefficients $a_2$ and $b_2$ were calculated in the same 
way using the ($I_{\mathrm{C,Rebull}}- I_{\mathrm{C,instr}}$) vs ($V_{\mathrm{instr}}-I_{\mathrm{C,instr}}$) 
plane. 

We typically used between 100 and 200 stars per chip and exposure time (minimum 42, maximum 392) for calculating 
the linear transformation coefficients. Thus even though many of the standard stars are variable this large number 
ensures a robust transformation. Obvious outliers were rejected before the calculation of the coefficients was done. 
Since the variable stars were in different phases of their brightness modulation for a given epoch the related data 
points scatter around the true transformation function. The uncertainties of the transformation coefficients results 
from this scatter. The median values of the slopes are $b_1=0.18$ and $b_2=0.10$. The typical uncertainties of the 
transformation coefficients are $\delta a_1=0.005$ and $\delta b_1=0.010$ for the first and $\delta a_2=0.015$ and 
$\delta b_2=0.010$ for the second equation. With this method we derived values for $I_{\mathrm{C}}$ at 7 different 
epochs between 9 Jan 2001 and 16 Jan 2001.

The final $I_{\mathrm{C}}$ magnitude for each star was calculated as the average of these 7 different $I_{\mathrm{C}}$ 
measurements. To improve the results we make use of the fact that we know the phase and amplitude of each star at these 
epochs from the relative photometric analysis described in Sect.~\ref{relphot}. Before the final averaging we subtracted
the relative magnitude ($m_\mathrm{rel}$) of the corresponding data point in the (mean subtracted) relative light curve from
the calculated $I_{\mathrm{C}}$ magnitude at each epoch. The mean of the resulting values is listed in Table~\ref{ergebnis}.

The errors for this final $I_{\mathrm{C}}$ magnitudes were estimated in two different ways. First, we calculated 
the expected 1$\sigma$ error $\delta I_{\mathrm{C,exp}}$ from the measured IRAF errors\footnote{The IRAF errors were 
corrected to somewhat higher values as described in section~\ref{iv}.} and the errors of the transformation coefficients. 
Second, we calculated the standard deviation $\sigma_7$ of the 7 independent $I_{\mathrm{C}}$ measurements. The final assigned 
error to the magnitude is the maximum of this two error estimations for each star: $\delta I_{\mathrm{C}} = 
max(\delta I_{\mathrm{C,exp}},\sigma_7)$.  

The colours in the Johnson $V$ and Cousins $R$, $I$ system were calculated in a similar way. The transformation from 
instrumental colour to the true colour was done using the transformation equations
\begin{eqnarray*}
    (R_{\mathrm{C}}-I_{\mathrm{C}}) = c_1 + d_1 \times (R_{\mathrm{C,instr}}-I_{\mathrm{C,instr}}) \hspace{1cm}
    \mathrm{or}
\end{eqnarray*}
\begin{eqnarray*}
   (V-I_{\mathrm{C}}) = c_1 + d_1 \times (V_{\mathrm{instr}}-I_{\mathrm{C,instr}}).
\end{eqnarray*}
Again, the transformation coefficients were calculated using a least square linear fit. The median values of the slopes 
are $d_1=-0.26$ and $d_2=-0.17$. With this method we derived ten values for ($V - I_{\mathrm{C}}$) and six for 
($R_{\mathrm{C}} - I_{\mathrm{C}}$). The final colours for each star were obtained by calculating the mean of these 
values for each colour. The errors were derived in the same way as described above.

\subsection{Final photometry database}\label{finalphot}

The absolute photometric calibration described in the previous section was done for each chip and exposure time separately. 
The objects that were measured on the same chip with two different exposure times were identified in order make sure
that they do not appear twice in the final catalogue. For stars with $I_{\mathrm{C}} \geq 16.0\,\mathrm{mag}$ the colours 
and magnitudes listed in Table~\ref{ergebnis} were taken from the 500\,sec exposures. The measurements for stars with 
$ 12.5\,\mathrm{mag} \leq I_{\mathrm{C}} \leq 16.0$ and with $10\,\mathrm{mag} \leq I_{\mathrm{C}}\leq 12.5\,\mathrm{mag}$ 
were taken from the 50\,esc and 5\,sec exposures, respectively.
Thus we list the measurements with the highest signal-to-noise for each star and avoid pixel saturation in images with the 
best seeing. The magnitude ranges used for our data analysis of the three different exposure times are also indicted in 
Fig.~\ref{meanerr}.

We compared the results of our photometry with those of other authors. In Table~\ref{phovgl} we report the mean offsets 
in the photometry between our study and those of other studies. We did not find any significant difference between our 
results and those reported by Rebull et al. (\cite{rebull2002}), Park et al. (\cite{park}), and Sung et al. (\cite{sung}).
The small differences between the studies can be explained by intrinsic stellar variability. However, there is a significant
offset in the photometry compared the the study of Flaccomio et al. (\cite{flaccomio}). The reason therefor could be 
different transmission curves of the used filters and a different average colour of the employed flux standards.

A small part of the final photometric catalogue is shown in Table~\ref{ergebnis} and the complete table that contains all 
10554 stars is available electronically. In Table~\ref{ergebnis} we also list the ($R_{\mathrm{C}} - H\alpha$) colour of 
the stars. Since no $H\alpha$ standard star measurements were available only the instrumental colour is given.
The limiting magnitudes in the different pass-bands for different signal-to-noise levels is given in Table~\ref{limitm}.

\begin{table*}[t]
\begin{minipage}[b][][t]{235mm}
  \centering
  \caption{Photometric data, the complete table available electronically. The column called ``vari'' lists the results of the 
	   time series analysis: pv refers to periodic and iv to irregular variables (see also the comments in the text). 
	   $\sigma$ is the standard deviation of the light curves from which outliers were rejected (see text).  
	   ptp is the estimated peak-to-peak variation in the light curves.} 
  \label{ergebnis}
  \begin{tabular*}{235mm}[b]{@{\extracolsep\fill}c@{\extracolsep\fill} c@{\extracolsep\fill} c@{\extracolsep\fill} c@{\extracolsep\fill}c @{\extracolsep\fill}c@{\extracolsep\fill}c @{\extracolsep\fill}c@{\extracolsep\fill}c @{\extracolsep\fill}c@{\extracolsep\fill}c @{\extracolsep\fill}c @{\extracolsep\fill}c @{\extracolsep\fill}c @{\extracolsep\fill}c @{\extracolsep\fill}c @{\extracolsep\fill}l} 
	\hline
	\hline
 Star & $\alpha (J2000)$&$\delta (J2000)$ & {$I_{\mathrm{C}}$}& {err} &{ $V-I_{\mathrm{C}}$} & {err} & {$R_{\mathrm{C}}-I_{\mathrm{C}}$} & {err} & {$R_{\mathrm{C}}-H\alpha$} & {err} &{vari}  & {$\sigma$} & {ptp}  & {SpT} & {name} & {cross id} \\ 
	\hline
6000 & 6:41:8.16 & 9:30:4.4  & 18.37 & 0.02 &      &      & 2.31 & 0.02 & -2.93 & 0.07 & iv     & 0.073 & 0.26 & & & \\ 
6001 & 6:41:8.19 & 9:36:56.8 & 16.91 & 0.01 & 4.06 & 0.06 & 2.06 & 0.01 & -3.32 & 0.02 & pv     & 0.053 & 0.17 & & & R3588 \\ 
6002 & 6:41:8.21 & 9:34:9.9  & 15.16 & 0.01 & 3.32 & 0.01 & 1.69 & 0.01 & -3.08 & 0.01 &        & 0.016 & 0.05 & & & R3591 \\ 
6003 & 6:41:8.21 & 9:38:30.9 & 16.56 & 0.01 & 3.83 & 0.08 & 1.93 & 0.01 & -2.88 & 0.04 & iv     & 0.074 & 0.27 & & & R3590 \\ 
6004 & 6:41:8.27 & 9:30:23.4 & 15.56 & 0.02 & 2.45 & 0.09 & 1.40 & 0.03 & -2.29 & 0.08 & iv     & 0.105 & 0.35 & & & \\ 
6005 & 6:41:8.28 & 9:38:14.7 & 13.10 & 0.01 & 0.73 & 0.01 & 0.35 & 0.01 & -3.38 & 0.01 &        & 0.010 & 0.04 & G5V$^1$ & & Y3607, R3592, MX234, W171 \\ 
6007 & 6:41:8.38 & 9:40:36.7 & 16.10 & 0.01 & 2.64 & 0.01 & 1.48 & 0.01 & -3.13 & 0.01 & pv?    & 0.015 & 0.06 & & & R3594 \\ 
6008 & 6:41:8.43 & 9:37:53.0 & 16.70 & 0.01 & 3.45 & 0.03 & 2.03 & 0.01 & -2.95 & 0.13 & pv     & 0.015 & 0.05 & & & R3595 \\ 
6010 & 6:41:8.52 & 9:32:51.5 & 19.13 & 0.02 & 2.34 & 0.03 & 1.10 & 0.01 & -2.54 & 0.02 & pv?/iv & 0.154 & 0.55 & & & \\ 
6011 & 6:41:8.53 & 9:23:6.6  & 20.54 & 0.06 & 0.90 & 0.04 & 0.44 & 0.04 &       &      &        & 0.126 & 0.52 & & & \\ 
6012 & 6:41:8.55 & 9:31:6.5  & 18.02 & 0.03 & 4.00 & 0.14 & 2.42 & 0.04 & -3.36 & 0.16 & pv     & 0.211 & 0.72 & & & \\ 
6013 & 6:41:8.56 & 9:42:52.1 & 14.57 & 0.02 & 1.91 & 0.01 & 1.01 & 0.01 & -3.23 & 0.01 & pv     & 0.062 & 0.18 & K7V$^1$ & & Y3618, R3598, S312, FX110, MX236, W173 \\ 
6014 & 6:41:8.57 & 9:30:4.2  & 17.20 & 0.01 & 3.35 & 0.02 & 1.97 & 0.02 & -2.74 & 0.01 & pv     & 0.016 & 0.05 & & & \\ 
6015 & 6:41:8.58 & 9:28:14.3 & 19.94 & 0.06 &-0.50 & 0.04 & 0.00 & 0.05 &       &      &        & 0.182 & 0.70 & & & \\ 
6016 & 6:41:8.63 & 9:53:47.3 & 17.74 & 0.01 & 3.06 & 0.03 & 1.95 & 0.01 & -2.95 & 0.02 & pv?    & 0.013 & 0.05 & & & R3600 \\ 
6018 & 6:41:8.65 & 9:39:11.9 & 20.31 & 0.08 &      &      & 2.29 & 0.15 &       &      &        & 0.056 & 0.21 & & & \\ 
6019 & 6:41:8.65 & 9:40:14.1 & 16.67 & 0.01 & 3.06 & 0.01 & 1.76 & 0.01 & -3.12 & 0.01 & pv     & 0.010 & 0.04 & & & R3599 \\ 
6020 & 6:41:8.70 & 9:53:25.3 & 20.28 & 0.06 &      &      &      &      &       &      &        & 0.068 & 0.25 & & & \\ 
6021 & 6:41:8.80 & 9:27:54.0 & 16.92 & 0.04 &-0.43 & 0.38 & 1.35 & 0.03 & -3.01 & 0.04 &        & 0.058 & 0.24 & & & \\ 
6022 & 6:41:8.81 & 9:23:43.9 & 12.91 & 0.02 & 1.83 & 0.02 & 0.98 & 0.02 & -3.13 & 0.02 & pv     & 0.073 & 0.26 & K4V$^1$ & & Y3650, R3602, P144, F381, FX111, MX240 \\ 
6023 & 6:41:8.84 & 9:48:48.6 & 21.58 & 0.19 &      &      &      &      &       &      &        & 0.331 & 1.18 & & & \\ 
6024 & 6:41:8.84 & 9:53:1.8  & 15.38 & 0.03 & 2.44 & 0.02 & 1.34 & 0.02 & -2.81 & 0.03 & pv     & 0.217 & 0.72 & & & \\ 
6025 & 6:41:8.85 & 9:45:28.0 & 18.83 & 0.08 &      &      & 2.90 & 0.04 & -2.02 & 0.82 &        & 0.021 & 0.07 & & & R3603 \\ 
6026 & 6:41:8.85 & 9:52:48.3 & 20.57 & 0.18 &      &      &      &      &       &      &        & 1.107 & 4.26 & & & \\ 
6027 & 6:41:8.86 & 9:46:1.6  & 12.31 & 0.02 & 0.71 & 0.01 & 0.38 & 0.01 & -3.36 & 0.02 &        & 0.014 & 0.05 & G5V$^1$ & & Y3628, R3606, MX237 \\ 
6028 & 6:41:8.89 & 9:29:45.7 & 19.50 & 0.11 & 1.28 & 0.05 & 0.71 & 0.07 &       &      &        & 0.238 & 1.07 & & & \\ 
6029 & 6:41:8.90 & 9:29:29.8 & 19.99 & 0.08 & 2.16 & 0.09 & 0.82 & 0.04 &       &      &        & 0.108 & 0.46 & & & \\ 
6030 & 6:41:8.92 & 9:29:22.6 & 19.59 & 0.06 & 2.45 & 0.14 & 1.12 & 0.05 &       &      &        & 0.151 & 0.70 & & & \\ 
 \hline
  \end{tabular*}
\end{minipage}
\end{table*}

\begin{table}
\centering
\caption{Limiting magnitudes for a signal-to-noise ratio of 20 and 50 of observations in each of the filters $V$, 
$R_{\mathrm{C}}$ and $I_{\mathrm{C}}$. Also listed is the estimated limit for completeness in each filter.} 
  \label{limitm}
  \begin{tabular}{ccc} 
	\hline
	\hline
	Filter & \multicolumn{2}{@{\extracolsep\fill}c}{Limiting Magnitude} \\
	       &  $20\sigma$ & $50\sigma$ \\
	\hline
	$V$              & $22.1$ & $21.1$ \\
	$R_{\mathrm{C}}$ & $21.5$ & $20.4$ \\
	$I_{\mathrm{C}}$ & $20.7$ & $19.6$ \\
	\hline
  \end{tabular}
\end{table}

\section{Time series analysis}\label{timeseries}

Our aim was to check all 10554 stars in the field for both periodic and irregular variability. For detecting these 
two types of variability we have used different techniques which are described in the following subsections. 

\subsection{Periodic variables}\label{pv}

We used two periodogram analysis techniques to search for significant periodicity in the light curves of each of the 
10554 monitored stars. Those stars which have fewer than 15 data points in their light curves were rejected from this 
analysis. $96\%$ of the remaining 10503 stars have 80 or more data points in their light curves. In the following 
we describe the two used periodogram algorithms separately.

\subsubsection{The Scargle Periodogram}

First we used the periodogram technique for unevenly sampled data described by Scargle (\cite{scargle}) and Horne \& 
Baliunas (\cite{horne}) to search for significant periodicity in all monitored stars. The algorithm calculates 
the normalised power $P_{\mathrm{N}}(\omega)$ for a given angular frequency $\omega=2\pi\nu$ and identifies the location
of the highest peak in the calculated periodogram of each star. Finally, the light curve of each star is phased to the
period according to the frequency of the highest peak. In order to decide whether there is a significant periodic 
signal of this period in the light curve, the height of this peak has to be related with a false alarm probability 
(FAP) which is the probability that a peak of this height is due simply to statistical variations, i.~e. noise.

The standard procedure for determining the FAP is to use simulated light curves created with Monte Carlo simulations.
It is necessary that the time sampling of the simulated light curves is identical with that of the measured light
curves. The simulated light curves represent non-variable stars and therefore contain only a intrinsic scatter in 
the photometry, i.~e. noise.
For each simulated light curve the periodogram has to be calculated and the power of the highest peak in that periodogram 
has to be determined. After various simulations (typically $\sim 10\,000$) the cumulative distribution of the power of
the highest peak is used to determine the FAP: The FAP of a given power $P_{\mathrm{N}}$ is set to be the fraction of 
simulated non-variable stars which have a highest peak power that exceeds $P_{\mathrm{N}}$. If we have simulated 
for example $10\,000$ light curves, the $1\%$ FAP power $P_{\mathrm{FAP=1\%}}$ is the power which was exceeded by the 
highest peak in $100$ simulations. 

The critical point in this procedure is the simulation of non-variable or at least non-periodic stars. The simplest 
approach to doing this is to assume that the data points are statistically independent of each other. 
This is only true if the typical time scale for intrinsic variations of the sources is not larger than the (typical) 
time difference of the data points. Strictly speaking this assumption is not valid for our sample since the significant 
time scale for variations in PMS stars is ${\sim\!1}~\mathrm{d}$ (e.~g. Herbst et al. \cite{herbst1994}) and in some 
nights our data points were obtained at closely spaced intervals of much shorter length ($\sim\!0.5~\mathrm{h}$). 
Therefore we calculated the FAP using different types of simulations. 

First we make the simplified assumption of uncorrelated data points and created light curves of pure noise using Monte 
Carlo simulations. The time sampling was chosen to be the same as in the real light curves. Gaussian distributed random
variables with zero mean and standard deviation $\sigma$ were assigned to these dates. We set $\sigma = 0.01\,\mathrm{mag}$ 
since the mean standard deviation in the light curves of the (non-variable) comparison stars we used for calculating the 
relative light curves is of the same order. Therefore we expect a standard deviation of $\sigma=0.01$ in the light curve 
of a non-variable star. After 10\,000 simulations we calculated a $1\%$ FAP power value of $P_{\mathrm{FAP=1\%}}=10.2$ 
from the cumulative distribution of the highest peak as described above.

To allow for correlations between the data points a second type of simulations was carried out. Instead of simulating 
pure Gaussian noise, the measured light curves of the monitored stars were used.
For the simulations we kept the time sampling the same as in the measured light curves. The relative magnitudes were 
randomly reassigned to the real dates. From the cumulative distribution of the maximum power we calculated a $1\%$ FAP 
power value of $P_{\mathrm{FAP=1\%}}=9.8$.

To be more conservative the highest power value of the two FAP simulations was used to define a cutoff level for the 
detection of periodic variables. Out of the 10554 analysed stars 1192 were brighter than $I_{\mathrm{C}}\leq\,19.5\mbox{mag}$ 
and had a peak power $P_{\mathrm{N}}\geq 10.2$. 
As an example we show in Fig.~\ref{phase_lc} the complete light curve and the resulting phased light curve we found for 
four stars. In Fig.~\ref{phase_pow} we show a sample of phased light curves at different power levels.

\begin{figure*}
  \centering
  \caption{Examples for phased light curves using the period $P_{Scargle}$ found by the Scargle periodogram technique. The 
	   peak power $P_N$ in the periodograms of these examples decreases from the upper left to the lower right panel. 
           The examples in the two top rows represent the highest power values ($P_N\simeq 40,\ldots,42$) we found, the 
	   examples in the two middle rows represent median power values ($P_N\simeq 28$) and the lower two rows represent
           the lowest power values we accepted ($P_N\simeq 11,\ldots,14$). The peak power $P_N$, the Period $P_{Scargle}$ 
	   and the star's identification number is given for each example. The error bar in the upper right corner
	   of each panel indicates the mean photometric error in the light curves.}
  \label{phase_pow}
\end{figure*}

\subsubsection{The CLEAN Periodogram}\label{clper}

The Scargle periodogram technique makes no attempt to account for the observational {\it window function} $W(\nu)$, 
i.~e. some of the peaks in the Scargle periodograms are normally a result of the data sampling. This effect is 
called aliasing and even the highest peak could be an artifact. The CLEAN periodogram technique by Roberts et al. 
(\cite{roberts}) tries to overcome this shortcoming of the Scargle periodogram. The observed power spectrum 
$|D(\nu)|^2$ is given by the convolution of the (complex) Fourier transformation $F(\nu)$ of the continuous signal 
from the star and the spectral window function $W(\nu)$ of the data set: 
\begin{equation}\label{convol}
D(\nu) \equiv F(\nu)\otimes W(\nu)
\end{equation}
(for a detailed description see Roberts et al. \cite{roberts} or Bailer-Jones \& Mundt \cite{caljm}).

Even if the spectral window function is known it is not possible to de-convolve $D(\nu)$ directly. Therefore Roberts et 
al. (\cite{roberts}) modified the CLEAN algorithm that is known from the reconstruction of two-dimensional images from 
interferometric data. Based on their publicly available Fortran code we have created a {\bf\it C} realization of the CLEAN 
algorithm. A description of this algorithm is given in Appendix \ref{appcl}. We calculated the CLEAN periodogram for each 
of the 10554 stars. After we had identified the highest peak in each of the periodograms the light curves of all stars were 
phased with the corresponding period $P_{CLEAN}$.

\subsubsection{The error in the measured periods}

The uncertainty in a measured period is set by two fundamental limits. First, it is related to the finite frequency 
resolution $\delta\nu$ of the power spectrum which makes it impossible to distinguish between closely separated periods 
and the uncertainty in a period is $\delta\nu /2$. For a discrete data set the resolution $\delta\nu$ is given by the 
width of the main peak of the window function $W(\nu)$. If the time sampling is not too non-uniform this is related to 
the total time span $T$ of the observations with $\delta\nu \simeq 1/T$ (Roberts et al. \cite{roberts}). Since for small 
errors $\delta\nu / \delta P \simeq \mathrm{d}\nu/\mathrm{d}p =1/P^2$ the error in the period is given by 
\begin{equation}
\delta P \simeq \frac{\delta\nu \, P^2}{2}.
\end{equation}
For larger errors (i.~e. larger periods) this is only a lower limit on the period error. Second, 
for very short periods when the finite time span between the data points 
is comparable to the period the uncertainty in a period is given by the sampling error which is related to the typical
spacing between the data points. Both of these fundamental limits leave their footprints in the main peak of the window 
function.

Since our time sampling is not very uniform (see Fig.~\ref{obsdates}) we determined the error of the periods directly 
from the width of this peak and for each star separately. The full width at half maximum of the main peak is typically 
(i.~e. for a light curve with 88 data points) $\nu_{FWHM} = 0.0260 1/\mathrm{day}$ which is $1.63$ times larger than 
what we would expect from the total duration of the observations. This can be explained by the non-uniform sampling. 
For a light curve with 88 data points we finally assigned an error to the period $P$ which is given by $\delta P =  
0.026 \times P^2$. For light curves with less data points the factor differs depending on the width of the main peak 
of the window function.

\subsubsection{Final period determination}

\begin{table*}[b]
  \caption{All periodic variables we found in our field. Listed are the identification number, the mean $I_{\mathrm{C}}$ 
	   magnitude, and the different periods detected by the Scargle periodogram technique ($P_{Sc}$), the CLEAN periodogram 
	   technique ($P_{CL}$) and by Kearns \& Herbst (\cite{KH}). Also shown is the period which we adopted in the end and the
	   corresponding error for this period. Those periodic variables that passed both of our PMS criteria are 
           marked with a {\it ``y''} in the column ``PMS member''. Those stars that failed at least one of the two tests are 
	   marked with a {\it ``n''} in the  appropriate column. In the latter case the superscript indicates which test the 
 	   star did not pass: a) rejected because of the location in $I_{\mathrm{C}}$ vs $(R_{\mathrm{C}}-I_{\mathrm{C}})$ diagram
	   and b) rejected because of the $(R_{\mathrm{C}}-H\alpha)$ colour. The full table is available electronically.} 
  \begin{minipage}{180mm}
  \begin{center}
  \label{pvtab}
  \begin{tabular}{@{\extracolsep\fill}rrrrrrrrc}	
   \hline
   \hline 
    Star & \multicolumn{1}{c}{$I_{\mathrm{C}}$} &  \multicolumn{1}{c}{(err)} &  \multicolumn{1}{c}{$P_{Sc}$} &  \multicolumn{1}{c}{$P_{CL}$} & \multicolumn{1}{c}{$P_{KH}$} &   \multicolumn{1}{c}{adopted $P$} &  \multicolumn{1}{c}{$\delta P$} &  \multicolumn{1}{c}{PMS} \\
   \hline
	  6001  & 16.91  &  0.01  &   6.01  &  5.98   &  &   6.01  &   0.50  &  n$^b$  \\
	  6008  & 16.70  &  0.01  &   1.32  &  1.32   &  &   1.32  &   0.02  &  y      \\
	  6012  & 18.02  &  0.03  &   4.35  &  4.33   &  &   4.35  &   0.26  &  n$^b$  \\
	  6013  & 14.57  &  0.02  &  10.48  & 10.46   &  &  10.48  &   1.43  &  y      \\
	  6014  & 17.20  &  0.01  &   1.59  &  1.59   &  &   1.59  &   0.03  &  y      \\
	  6019  & 16.67  &  0.01  &   0.65  &  0.64   &  &   0.65  &   0.01  &  y      \\
	  6022  & 12.91  &  0.02  &   1.73  &  1.73   &  &   1.73  &   0.05  &  y      \\
	  6024  & 15.38  &  0.03  &   9.71  &  9.81   &  &   9.71  &   1.23  &  y      \\
	  6031  & 13.85  &  0.01  &   3.92  &  3.90   &  &   3.92  &   0.20  &  y      \\
	  6032  & 14.72  &  0.01  &   0.80  &  0.80   &  &   0.80  &   0.01  &  y      \\
	  6039  & 14.44  &  0.01  &   5.83  &  5.76   &  &   5.83  &   0.44  &  y      \\
	  6042  & 16.91  &  0.01  &   0.54  &  0.54   &  &   0.54  &   0.00  &  y      \\
	  6043  & 15.01  &  0.01  &  10.77  & 10.64   &  &  10.77  &   1.51  &  y      \\
	  6045  & 17.19  &  0.01  &   1.75  &  1.76   &  &   1.75  &   0.04  &  y      \\
	  6055  & 13.64  &  0.01  &   5.92  &  5.98   &  &   5.92  &   0.47  &  y      \\
	  6063  & 15.33  &  0.01  &   3.38  &  3.38   &  &   3.38  &   0.15  &  y      \\
	  6064  & 16.13  &  0.01  &   2.18  &  2.17   &  &   2.18  &   0.07  &  y      \\
	  6067  & 12.78  &  0.01  &   2.93  &  2.92   &  &   2.93  &   0.14  &  y      \\
	  6077  & 19.77  &  0.03  &   2.40  &  2.37   &  &   2.40  &   0.08  &  y      \\
	  6079  & 15.83  &  0.01  &   5.22  &  5.23   &  &   5.22  &   0.35  &  y      \\
	  6081  & 15.07  &  0.01  &   0.68  &  0.68   &  &   0.68  &   0.01  &  y      \\
	  6093  & 16.53  &  0.01  &   0.96  &  0.96   &  &   0.96  &   0.01  &  n$^{a,b}$\\
	  6101  & 18.00  &  0.01  &   0.46  &  0.46   &  &   0.46  &   0.00  &  n$^a$  \\
	  6102  & 15.15  &  0.01  &   9.04  &  9.10   &  &   9.04  &   1.06  &  y      \\
	  6115  & 17.08  &  0.01  &   0.71  &  0.71   &  &   0.71  &   0.01  &  y      \\
  \hline
  \end{tabular}
  \end{center}
  \end{minipage}
\end{table*}

For all of the 1192 stars that have a normalised peak power in the Scargle periodogram of $P_N \geq 10.2$ (i.~e. FAP 
$\leq 1\%$) both the light curve and the light curves phased with the periods found by both periodogram techniques were 
checked by eye and obvious wrong detections (e. g. due to light contamination from a close-by neighbour) were rejected. 

The highest frequency which may be recovered from a sample with data points spaced at intervals of $\Delta$ is the Nyquist 
frequency $\nu_{\mathrm{N}} =1/(2\Delta)$ (Roberts et al. \cite{roberts}). Using $\Delta=0.107\,\mbox{day}$ which is 
the median time difference of the data points in a light curve we therefore set a lower limit on the periods and accepted 
only periods with $P\geq 1/\nu_{\mathrm{N}} = 2\Delta = 0.21\,\mbox{day}$.

In total we found 543 periodic variable stars. These stars are marked in Table~\ref{ergebnis} with a {\it pv} in the
column ``{\it vari}''. For 520 ($95.8\%$) of these periodic variables the periods determined with the two different 
periodogram techniques agree within the estimated errors. In all cases where the periods determined with both methods 
did not agree the two periods were beat period of each other. For those stars we assigned one of the two periods to the 
star. We always choose the period for which the scatter in the phased light curve was lowest. For 17 of these 23 stars 
we assigned the CLEAN period but we kept the Scargle Period in six cases.

In addition all 543 periodic variables we found are listed in a separate Table~\ref{pvtab}. In this table we give 
the periods we found with both the Scargle and the CLEAN periodogram techniques, the period we finally assigned to each 
star, and the estimated errors for this period. We also list the periods measured by Kearns \& Herbst (\cite{KH}), if 
available. Out of the 543 stars 501 ($91.9\%$) have a peak power in the Scargle periodogram of $P_N\geq 12.40$ which 
corresponds to a FAP of $0.1\%$.

The locations of the periodic variables in our observed field are shown in Fig.~\ref{positionen}. Their distribution 
show that there are two concentrations of young stars in NGC\,2264 which we call NGC\,2264\,N (north) and NGC\,2264\,S
(south). These concentrations are already known in the literature: Sagar et al. \cite{sagar1988} report these two 
points of maximum stellar density of cluster members. This clustering agrees with the observed distribution of molecular
gas, H$\alpha$-emission stars and early type stars in the cluster (Mathieu \cite{mathieu1986} and references therein).
The stars located in these two concentrations will be used in Sect.~\ref{selCMD} to determine the region in the 
colour-magnitude diagram where the PMS stars are located.

Out of the stars which we rejected after our visual inspection another 136 were classified as possible periodic variables. 
These stars are marked with a {\it pv?} in the column ``{\it vari}'' of Table~\ref{ergebnis} but are not listed in 
Table~\ref{pvtab}, because they are not used for any further analysis (see Paper II).

\subsection{Irregular variables}\label{iv}

Since many of the PMS stars in our study show non-periodic brightness modulations with a peak-to-peak variation up to 
$1\,\mathrm{mag}$ we defined any variable star for which no period could be detected as an irregular variable.
 
We searched for this irregular variability in the light curves of all stars that are not periodically variable in the 
magnitude range of $I_{\mathrm{C}}\leq 19.5$ (which corresponds to $S/N \geq 50$) that have at least 20 data points. 
Stars with close neighbours were rejected from this sample because our fixed-size aperture could lead to seeing-dependent 
overlapping with a (bright) neighbour, and hence imitate variability.
Since the impact of a faint neighbour is negligible we used a rejection criteria that is magnitude dependent. If 
two stars were separated by less than 2\farcs 5 from each other, both of them were rejected from the 
analysis if the magnitude difference between the two objects was less than 2.0\,mag. If the magnitude difference 
of such two stars is larger than 2.0\,mag only the fainter star was removed and the brighter stars was kept.

In total we analysed 5927 non-periodic stars. $90.6\%$ of these stars have more than 70 data points in the light curve. 
For detecting the irregular variables we used a $\chi^2$ test, in which we calculate the probability that the deviations in 
the light curve are consistent with the photometric errors, i.e the probability that the star is not variable. Therefore we 
evaluate 
\begin{equation}\label{cisqeq}
\chi^2 = \sum_{j=1}^N \left( \frac{m_{\mathrm{rel}}(j)}{\delta m_{\mathrm{rel}}(j)}\right)^2
\end{equation}
where $m_{\mathrm{rel}}(j)$ is the relative magnitude and $\delta m_{\mathrm{rel}}(j)$ is the error of the $j$th data point 
in the light curve of a star. 
The probability that the light curve of a non-variable star with $N$ data points results in a value for chi-square that 
exceeds the measured value $\chi ^2$ is given by \footnote{Since the time series has been mean subtracted the number of 
degrees of freedom is N-1.} (Press et al. \cite{press})
\begin{equation}
Q(\chi^2|N)=\frac{\Gamma\left(\frac{N-1}{2},\frac{\chi^2}{2}\right)}{\Gamma\left(\frac{N-1}{2}\right)}\equiv 
\frac{\int_{\chi^2/2}^\infty t^{(N-3)/2} e^{-t} dt}{\int_0^\infty t^{(N-3)/2}  e^{-t} dt}.
\end{equation}
The probability $P_{\mathrm{vari}}$ that the star is variable is therefore $P_{\mathrm{vari}}=1-Q(\chi ^2|N)$. 

We note that the $\chi^2$ test is very sensitive to a over- or underestimation of the errors. For example if the 
error in the single measurements is underestimated by a factor of 2 the value for $\chi^2$ is overestimated by a factor of 
4 and we therefore deduce an overestimated probability $P_{vari}$ that the star is variable. In order to check whether the 
errors in the light curves are over- or underestimated we made a comparison of the scatter $\sigma$ and the average of the 
errors $\overline{\delta m_{\mathrm{rel}}} = (1/N) \sum_{j=1}^N \delta m_{\mathrm{rel}}$ in the light curves, where $\delta 
m_{\mathrm{rel}}$ is given by Eq.~(4) in Bailer-Jones \& Mundt (\cite{caljm}). 
\begin{figure}
  \resizebox{\hsize}{!}{\includegraphics{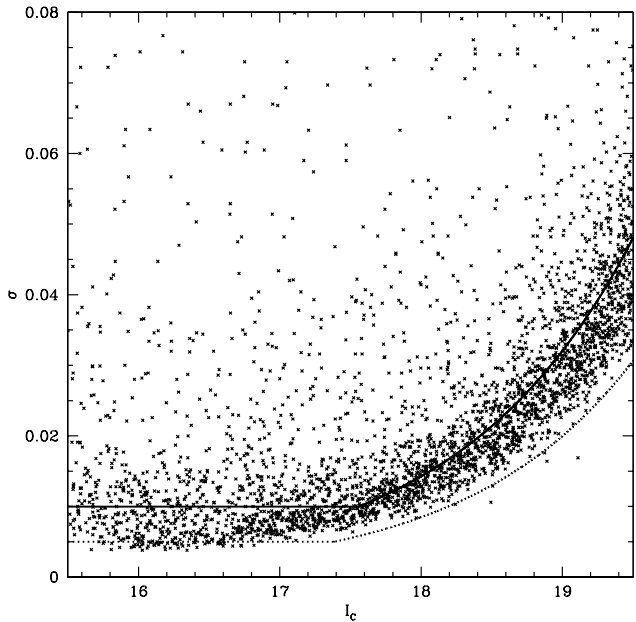}}
    \caption{Standard deviation $\sigma$ of all stars measured with the 500\,sec exposures as a function of magnitude. 
	     The solid line represents a fit ($S(I_{\mathrm{C}})$) to median values of $\sigma$. The dotted line shows the  
             fit ($M(I_{\mathrm{C}})$) to the median of the mean photometric error $\overline{\delta m_{\mathrm{rel}}}$ 
             displayed in Fig.~\ref{meanerr}. The medians for both quantities were calculated in equally spaced magnitude 
             bins.}
    \label{sigerr}
\end{figure}
In Fig. \ref{sigerr} we show the scatter in the light curve of each star measured in the 500\,sec exposures. The dotted 
line is a fit  $M(I_\mathrm {C})$ to the median of the mean photometric errors of the 500\,sec exposures shown in the lower 
panel of Fig.~\ref{meanerr}. The solid line represents the fit $S(I_\mathrm {C})$ of the median of the standard deviation 
$\sigma$ in the light curves. The median values for both the standard deviation $\sigma$ and the mean error $\overline{\delta 
m_{\mathrm{rel}}}$  were calculated in different $I_\mathrm {C}$ magnitude bins. In both cases we fitted an exponential 
function in the fainter regime and a constant in the brighter regime. 

It is evident that the estimated errors $\delta m_{\mathrm{rel}}$ are systematically underestimated for the 500\,sec 
exposures and the same is the case for the 5\,sec and 50\,sec. Therefore we corrected the errors measured for each star by
calculating the ratio of the two functions for each of the three exposure times to give a correction function 
$C(I_\mathrm{C})=S(I_\mathrm {C})/M(I_\mathrm {C})$. 
The error of each data point in the light curves of a given star was corrected using the equation 
\begin{equation}
\delta m_{\mathrm{rel,cor}}(j) = C(I_\mathrm {C})\, \delta m_{\mathrm{rel}}(j)
\end{equation}
where we used the mean magnitude for $I_\mathrm{C}$ of the star calculated as described in Sect.~\ref{absphot}. By correcting
the errors in this way we ensure that the relative distribution of the errors in a given light curve is conserved.

We note that we used a conservative fit for the standard deviation in the sense that the standard deviation may be 
overestimated by our fit (see Fig.~\ref{sigerr}). Since the same is the case for the fit $M(I_\mathrm {C})$ of the mean 
errors this effect is minimised (but maybe not completely canceled) for the ratio $C(I_\mathrm{C})=S(I_\mathrm {C})
/M(I_\mathrm {C})$. However, a conservative estimate of the true error reduces the false detections of the 
$\chi^2$-test. On the other hand the test is less sensitive for low amplitude variations. 

From Eq.~\ref{cisqeq} it is clear that a single outlier in the light curve (e. g. caused by a cosmic ray) could produce 
a very high value for the $\chi ^2$. Since we are looking only for persistent variability we have applied a sigma clipping 
algorithm to the light curves before the $\chi ^2$-test was performed. This algorithm uses the standard deviation 
$\sigma$ of a given light curve and removes all data points from this light curve that are located more than $2.5\sigma$ 
above or below zero (note that the mean value was subtracted from the light curves).

After all stars had been analysed we found that many of the detected variable stars are located in the corners of our 
field. The light curves of these stars looked very similar and were well correlated with the changing seeing.
A possible explanation for this is that the pixel scale of the WFI is different in the corners of the field. A similar 
result was recently found by Koch et al. (private communication) for the WFI. This leads to variations in the photometry 
of the stars located in the corners because the flux in the aperture of these stars changes with the seeing in a 
different way from the flux in the apertures of the comparison stars (the latter are not located in the corners, see 
Sect.~\ref{relphot}). Therefore non-variable stars can mimic variability. 
In order to avoid this problem resulting from the variable pixel 
scale we performed the relative photometry again in exactly the same way as described in Sect.~\ref{relphot} but we used 
this time a significantly larger aperture diameter of 20\,pixels (4\farcs 76) instead of 8\,pixels. In this way we made sure 
that much more of the star's light falls into the aperture and the measurement is much less dependent on the seeing, i.~e. 
the photometric errors dominate the changes due to seeing variations. The errors in the relative magnitudes were corrected 
in the same way as described above. 

In order to decide whether any star in the field is variable we used the results of the $\chi^2$-tests based on the 
measurements with both aperture sizes. Only the stars that have a probability of $P_{vari} \geq 99.9\%$ in {\it both} 
tests were assumed to be irregular variable. We used both aperture sizes and not only the measurements with the bigger 
aperture because with the large aperture we have increased the probability that the measurement of the stellar brightness 
is contaminated by the light of a close-by neighbour and the star therefore can mimic variability with the changing seeing. 
This effect is minimised for the photometry with the smaller aperture radius. On the other hand stars that do mimic 
variability because of the variable pixel scale do not pass the $\chi^2$-tests based on the measurements with a bigger 
aperture radius. Hence, a star that passes both tests is not affected by the variable pixel scale and light contamination 
from a close-by neighbour and therefore the variability is intrinsic to the star.

\begin{figure}
  \resizebox{\hsize}{!}{\includegraphics{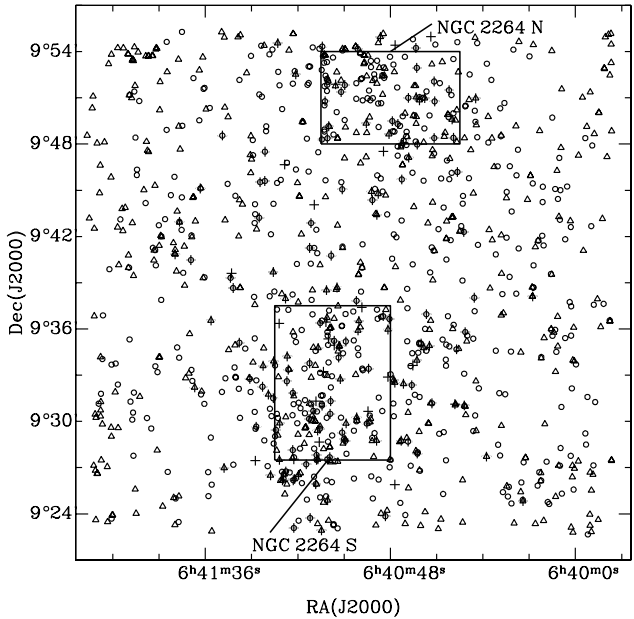}}
    \caption{The locations of all 543 periodic ({\large $\circ$}) and 484 irregular ($\triangle$) variables from our survey
	     (including non-PMS stars). Those variable stars which were previously known as PMS stars are indicated by a 
	     vertical line (\,{\large $\circ\!\!\!\!\mid\, ,\: \triangle\!\!\!\!\mid$ } respectively) while previously
	     known PMS members which are not variable according to our investigations are marked by a cross ({\large$ +$}). 
	     The two regions NGC\,2264\,N \& S with the highest concentration of variables are marked by boxes.}
    \label{positionen}
\end{figure}

In total we found 484 irregular variables out of the 5927 stars we analysed. The irregular variables are listed in column 
``{\it vari}'' of Table~\ref{ergebnis} as {\it iv}. The spatial positions of the irregular variables are shown in 
Fig.~\ref{positionen}. We remind the reader that irregular variables are defined here as variables with $P_{vari} \geq 
99.9\%$ for which we could not find any significant period in our periodograms.

With the $\chi^2$ test we are able to detect variability for stars brighter than 
$I_{\mathrm{C}}=16.0\,\mbox{mag}$ if the standard deviation $\sigma$ in the light 
curves is larger than 0.02\,mag. Since we used images with 500\,sec exposure 
time for stars fainter than $I_{\mathrm{C}}\leq 16.0\,\mbox{mag}$, the sensitivity 
of the $\chi^2$ test is somewhat better for stars with $16.0\,\mbox{mag} \leq 
I_{\mathrm{C}}\leq 17.25\,\mbox{mag}$. In this magnitude range we are able to detect 
variability for stars with $\sigma \geq 0.015\,\mbox{mag}$. For stars with 
$I_{\mathrm{C}}=18\,\mbox{mag}$ we are able to detect variability if 
$\sigma \geq 0.03\,\mbox{mag}$.

Only $48.3\%$ of the periodic variables are variable according to our $\chi ^2$-test. It is not surprising that 
not all periodic variables are variable according to this test because the periodogram analysis is much more sensitive 
to low amplitude variations than the $\chi^2$ test (see Sect.~\ref{natvari}). Furthermore we have adopted relatively 
conservative errors for our $\chi^2$ test (see above). This decreases the sensitivity to small amplitude variations.

We note that the different pixel scale in the corners of the field has only weak or no effect on the results of our 
periodic variability study. Since the seeing changes randomly the amplitudes of this random variation contributes to all 
powers in the power spectrum and not only to a single peak. Only $\sim 5-10$ periodic variables are located in the 
affected regions of the field. However, the variations in the corners of the field might have caused us to miss detecting 
a small number periodic variables.

\section{PMS Membership}\label{member}

The periodic and irregular variables we found in our field are not necessarily all PMS stars, although 
variability is probably one of the best indicators for youthfulness. Nevertheless there will be a certain 
degree of contamination by non-PMS stars if one selects candidates on the basis of variability. Therefore 
we have to disentangle variable PMS stars from variable background or foreground stars. This was done in 
two steps, namely using first the $I_\mathrm{C}$ vs ($R_\mathrm{C}-I_\mathrm{C}$) colour-magnitude diagram 
(CMD) and second the ($R_\mathrm{C}-H\alpha$) vs ($R_\mathrm{C}-I_\mathrm{C}$) colour-colour diagram. In 
the following subsections we describe the two discrimination procedures in detail. Variable stars which 
pass both of these tests will be likely PMS members of the cluster and for sake of simplicity will therefore  
be called PMS stars in the subsequent discussion. For a full confirmation of the PMS nature of these stars  
additional observations are necessary, like measurements of the Li\,I\,$\lambda 6707$-line equivalent width.  
We expect, however, that only a minor fraction of the stars which passed both of the tests are non-PMS stars.

\subsection{PMS test I: The colour-magnitude diagram and determination of the PMS region}\label{selCMD}

In a CMD the PMS stars are located in a region above the main sequence (MS). This is due to their larger stellar radii 
and correspondingly larger brightness compared to MS stars of the same spectral type or colour. 
In order to determine the borders of the PMS region we placed a well defined sample of PMS stars in the CMD. This sample 
was selected in such a way that it is contaminated as little as possible by background and foreground stars. It consists 
of two subsamples: 1) previously known cluster members selected from the catalogues of Park, Sung \& Bessel (\cite{park}) 
and Sung, Bessel \& Lee (\cite{sung}) and 2) a well selected subsample of the newly found periodic and irregular variables. 
The second subsample consists of our newly found periodic and irregular variables located in the two dense concentrations 
of variables in the cluster. These two concentrations are called NGC\,2264\,N~\&~S and are 
marked in Fig.~\ref{positionen}. The two regions are a part of the so called ``on cloud region'' defined by Rebull et al. 
(\cite{rebull2002}) and as already mentioned above are known as regions of maximum stellar density in the cluster 
(Mathieu \cite{mathieu1986}, Sagar et al. \cite{sagar1988}).

Determining the borders of the PMS region in the CMD only from the variable stars in NGC\,2264\,N~\&~S has the advantage 
of producing a maximum fraction of PMS stars relative to any background stars. This is due to high extinction of the dust located 
behind the cluster stars. This is shown by Fig.~\ref{pos_back} which also illustrates that in particular for NGC\,2264\,S 
a very large background extinction is present. Since we selected only {\it variable} stars located in these regions the 
contamination of the this subsample with (non-variable) MS foreground stars is also negligible. Therefore the selected 
variable stars in the two regions NGC\,2264\,S and N are most likely PMS cluster members. 

\begin{figure}
  \resizebox{\hsize}{!}{\includegraphics{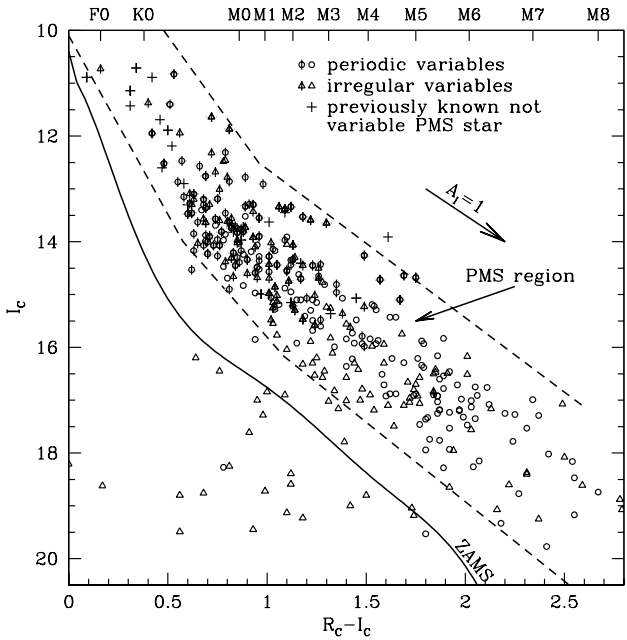}}
    \caption{Determination of the PMS region in the CMD. Only the variables in NGC\,2264\,S and N (see Fig.\,\ref{positionen})
	     and all previously known PMS stars in our whole \object{NGC\,2264} field are plotted. The symbols are 
	     the same as in Fig.~\ref{positionen}. The two dashed lines indicate the lower and upper borders of the PMS 
	     region.
     \label{defregion}}
\end{figure}

The CMD of all variable stars in NGC\,2264\,N~\&~S and all previously known PMS stars in our whole NGC\,2264 field 
is shown in Fig.~\ref{defregion}. The spectral types marked at the top of the panel were determined by using 
photometric measurements of ZAMS stars with known spectral type. For stars with spectral types earlier than M0 
mesurements reported by Johnson (\cite{johnson1966}) were used. For spectral types later than M0 measurements by 
Leggett (\cite{leggett1992}) and Kirkpatrick \& McCarthy (\cite{kirkpatrick1994}) were used. If necessary, 
transformation into the Cousins system were performed by applying the transformation equations by Bessel 
(\cite{bessel1983}). For the spectral types zero reddening is assumed.
From this diagram we defined the PMS region in the CMD. The lower and upper borders of this region are indicated by the 
dashed lines.

The defined PMS region will obviously eliminate MS stars which are at the same distance as NGC\,2264 or further 
away. Such stars are located in the CMD at or below the ZAMS indicated in Fig.~\ref{defregion} and are therefore 
well outside the PMS region. Because of their larger apparent $I_{\mathrm{C}}$ magnitude foreground MS stars with a 
distance between $\sim300\,\mathrm{pc}$ and $\sim600\,\mathrm{pc}$ could be located in the PMS region. However, as 
already mentioned above the contamination with these stars is probable negligible.

On the other hand it is {\it not possible} to discriminate between variable PMS stars and variable background 
giants using the PMS region in the CMD if the latter are at (large) distances which results in apparent magnitudes 
that shift them into the PMS region. K or M giants are typically between 3 and 10 magnitudes brighter than PMS 
stars at the same distance and the same spectral type. Therefore background giants with a corresponding smaller 
apparent $I_{\mathrm{C}}$ magnitude could be located in the PMS region of Fig.~\ref{defregion}. Since the reddening 
vector is nearly parallel to the borders of the PMS region even a highly reddened giant will stay in the PMS region. 
In addition several types of giants are variable (e.~g. RR\,Lyrae stars) and therefore it is necessary to eliminate 
these stars from our sample of variable stars using a different selection criterion. This will be done in 
Sect.~\ref{selCCD}

Of the 543 periodic variables we found in our whole NGC\,2264 field, 451 stars were located in this PMS region 
and only these stars were used for the further analysis. These numbers together with the few (only 3) periodic variables 
outside the PMS region in Fig.~\ref{defregion} confirm that the regions NGC\,2264\,S~\&~N are very little contaminated 
with foreground or background MS stars compared to the rest of our field. That furthermore indicates that the above 
described procedure to define the PMS region in the CMD provides relatively reliable borders.

\begin{figure}
    \caption{The ($R_\mathrm{C}-H\alpha$) vs ($R_\mathrm{C}-I_\mathrm{C}$) colour-colour diagram for all stars which passed the 
	     first PMS test. Circles and triangles represent periodic and irregular variables respectively. Filled symbols 
	     indicate the stars which passed both PMS tests. Variables which were previously known as PMS are indicated by a 
	     vertical line. Non-variable previously known PMS stars are indicated by a cross. The solid line represents the 
	     fit of the locus of PMS and MS stars. Variable stars below the dashed line (open symbols) were rejected from 
	     the analysis because they are probable background giants.}
    \label{halphafig}
\end{figure}
\begin{table*}
\centering
\caption{The relation between ($R_\mathrm{C}-I_\mathrm{C}$) and ($R_\mathrm{C}-H\alpha$) colours for PMS/MS locus 
	 stars. The  values for ($R_\mathrm{C}-H\alpha$) are calculated in bins of 0.05~magnitude width. $\sigma$ is the
	 standard deviation of the ($R_\mathrm{C}-H\alpha$) colours in each bin.}
  \label{MSrel}
  \begin{tabular}{ccc|ccc|ccc|ccc} 
	\hline
	\hline
	$(R_\mathrm{C}-I_\mathrm{C})$ & ($R_\mathrm{C}-H\alpha$) & $\sigma$ &
	$(R_\mathrm{C}-I_\mathrm{C})$ & ($R_\mathrm{C}-H\alpha$) & $\sigma$ &
	$(R_\mathrm{C}-I_\mathrm{C})$ & ($R_\mathrm{C}-H\alpha$) & $\sigma$ &
	$(R_\mathrm{C}-I_\mathrm{C})$ & ($R_\mathrm{C}-H\alpha$) & $\sigma$ \\
	\hline
 0.38 & -3.36 &  0.05 &    0.83 & -3.22 & 0.06 &   1.28 & -3.10 & 0.08 &   1.73 & -3.02 &  0.12\\
 0.43 & -3.33 &  0.04 &    0.88 & -3.24 & 0.05 &   1.33 & -3.10 & 0.07 &   1.78 & -3.03 &  0.13\\
 0.48 & -3.33 &  0.05 &    0.93 & -3.22 & 0.06 &   1.38 & -3.08 & 0.08 &   1.83 & -2.99 &  0.15\\ 
 0.53 & -3.34 &  0.05 &    0.98 & -3.22 & 0.05 &   1.43 & -3.09 & 0.07 &   1.88 & -3.02 &  0.11\\
 0.58 & -3.30 &  0.04 &    1.03 & -3.20 & 0.06 &   1.48 & -3.07 & 0.10 &   1.93 & -3.02 &  0.14\\
 0.63 & -3.30 &  0.06 &    1.08 & -3.17 & 0.06 &   1.53 & -3.07 & 0.10 &   1.98 & -2.92 &  0.15\\
 0.68 & -3.30 &  0.05 &    1.13 & -3.16 & 0.07 &   1.58 & -3.06 & 0.10 &   2.03 & -3.03 &  0.11\\
 0.73 & -3.26 &  0.04 &    1.18 & -3.13 & 0.07 &   1.63 & -3.04 & 0.10 &   2.08 & -2.89 &  0.13\\
 0.78 & -3.24 &  0.04 &    1.23 & -3.12 & 0.08 &   1.68 & -3.05 & 0.10 &   2.13 & -2.87 &  0.20\\
	\hline
  \end{tabular}
\end{table*}

\subsection{PMS test II: The ($R_\mathrm{C}-H\alpha$) vs ($R_\mathrm{C}-I_\mathrm{C}$) colour-colour diagram}\label{selCCD}

\begin{figure}
  \resizebox{\hsize}{!}{\includegraphics{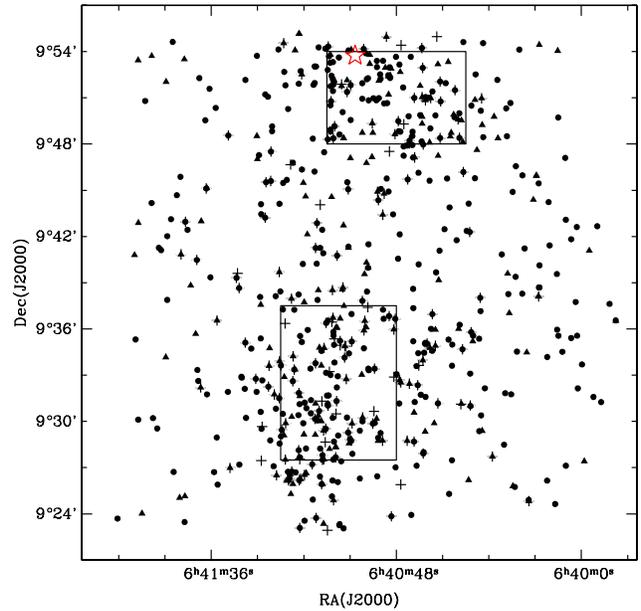}}
    \caption{Locations of all PMS variables from our survey which passed both tests including non-variable previously known
	     PMS members. The symbols are the same as in Fig.\,\ref{halphafig}. Also shown (indicated by the two boxes) are 
	     the two regions NGC\,2264\,N \& S which we used to determine the PMS region in the CMD. The star in the northern
	     box indicates the location of the bright star S Mon (V=4.7\,mag).} 
    \label{location_pv_iv}
\end{figure}
\begin{figure}
  \resizebox{\hsize}{!}{
}
    \caption{Locations of all stars in the field that {\it failed} at least one of our PMS membership criteria 
	     (see Sect.~\ref{member}). We also show the two regions which we used for determination of the PMS region in 
	     the CMD.} 
    \label{pos_back}
\end{figure}
\begin{figure}
    \caption{The $I_{\mathrm{C}}$ vs ($R_{\mathrm{C}}-I_{\mathrm{C}}$) colour-magnitude diagram for different samples of 
	     stars. In (a) we show all 543 periodic variables we found. The 405 periodic variables that passed our two 
	     PMS criteria are marked with filled circles while those periodic variables which did not pass at least 
	     one test are marked by open circles. The PMS zone is indicated by the dashed lines. The solid line represents 
	     the ZAMS sequence. Panel (b) shows the CMD for the all 484 variable stars. The 184 irregular variables which  
	     fulfilled both of our PMS criteria are indicated by filled triangles. The 300 variable stars which failed  
	     at least one test are shown by open triangles.
	     In panel (c) we show the CMD of the previously known PMS stars in our whole NGC\,2264 field.} 
    \label{FHD}
\end{figure}

As outlined in the previous section it is essential to eliminate background giants from the sample of variables we 
found. Therefore we used the location of the stars in the ($R_\mathrm{C}- H\alpha$) vs ($R_\mathrm{C}-I_\mathrm{C}$) 
colour-colour diagram (Fig.~\ref{halphafig}) as a second additional selection criterion for classifying a star 
as a PMS star; i.~e. all stars which passed the first test have also to pass the second one to be classified as a PMS 
star. Many PMS stars, in particular CTTSs, show large $H\alpha$ emission. Thus their ($R_\mathrm{C}-H\alpha$) colour 
is larger than the ($R_\mathrm{C}-H\alpha$) colour of a MS star of the same spectral type. There also exists WTTSs or 
``naked'' TTSs with weak or no detectable $H\alpha$ emission (e.g. see Appenzeller \& Mundt, 1989). The locus of these 
latter stars is similar to the MS in the ($R_\mathrm{C}- H\alpha$) vs ($R_\mathrm{C}-I_\mathrm{C}$) colour-colour 
diagram.

On the other hand giants with spectral types later than $\sim$K3 have smaller ($R_\mathrm{C}- H\alpha$) colours 
than MS stars of the same ($R_\mathrm{C}-I_\mathrm{C}$) colour. The differences in ($R_\mathrm{C}- H\alpha$) 
between giants and PMS/MS stars result from stronger molecular bands in the giants' spectra causing different 
spectral energy distributions in the $R_{\mathrm{C}}$ band. In addition giants are typically at larger distances 
and are therefore highly reddened. Since reddening affects mainly the ($R_\mathrm{C}-I_\mathrm{C}$) colour they 
are shifted further away from the MS (i.~e. to the right) with increasing distance. This behaviour was 
confirmed by a simulated ($R_\mathrm{C}- H\alpha$) vs ($R_\mathrm{C}-I_\mathrm{C}$) colour-colour diagram 
with the help of standard star spectra of different spectral type and luminosity (i.~e. by multiplying the  
filter transmission curves of the WFI with the spectral energy distribution of standard stars spectra).

Therefore it is possible to discriminate between (background) giants and PMS stars using their location in the 
($R_\mathrm{C}-H\alpha$) vs ($R_\mathrm{C}-I_\mathrm{C}$) colour-colour diagram. We determined the median 
($R_\mathrm{C}-H\alpha$) colour as a function of $(R_\mathrm{C}-I_\mathrm{C})$ for the locus of PMS/MS stars using 
the data shown in Fig.~\ref{halphafig}. In Table~\ref{MSrel} 
we list the median ($R_\mathrm{C}-H\alpha$) colour as a function of $(R_\mathrm{C}-I_\mathrm{C})$ for the stars on the PMS/MS 
locus. These values were derived in different ($R_\mathrm{C}-I_\mathrm{C}$) colour bins of 0.05 magnitude width. We have fitted 
a quadratic function to these medians and obtained a locus relation of the form 
\begin{equation}\label{rhams}
(R_\mathrm{C}-H\alpha)_{\mathrm{locus}}= - 0.06\times(R_\mathrm{C}-I_\mathrm{C})^2 + 0.38\times(R_\mathrm{C}-I_\mathrm{C}) - 3.50.
\end{equation} 
This relation is shown in Fig.~\ref{halphafig} as the solid line. As we will show in Sect.~\ref{sec_haindex}),
the vertical distance of a star from this line measures the strength of the star's H$\alpha$ emission.

In order to discriminate between PMS stars and background giants we defined a lower discrimination level shown in 
Fig.~\ref{halphafig} as a dashed line. This line is given by $(R_\mathrm{C}-H\alpha)_{\mathrm{low}} - 1.65\times\delta$, 
where $\delta$ is a fit of the standard deviation in $(R_\mathrm{C}-H\alpha$) in each $(R_\mathrm{C}-I_\mathrm{C})$ bin and 
is given by $\delta = e^{0.60\,(R_\mathrm{C}-I_\mathrm{C})-3.42}$. Assuming that the locus of stars in each bin are Gaussian 
distributed $95\%$ of the locus stars are located above the lower discrimination level which therefore represents a lower
envelope of the locus. Any variable star (periodic or irregular) which passed the first selection criterion described in 
Sect.~\ref{selCMD} and that is located above this level was kept for the final analysis while stars below this level were 
rejected.

From the 451 periodic variables that passed the first selection criterion another 46 were rejected because of their 
$(R_\mathrm{C}-H\alpha)$ colour. We finally have a total of 405 periodic and 184 irregular variables (with $P_{vari} \geq 
99.9\%$) which are all very likely PMS members of the NGC\,2264 star forming region. Hence, we call these stars variable
PMS members. The periodic variable PMS members are separately marked in Table~\ref{pvtab}.

\subsection{The spatial distribution of PMS members}\label{pvdatabase}

In Fig.~\ref{location_pv_iv} and Fig.~\ref{pos_back} we show the spatial distribution of all PMS stars and the positions 
of non-cluster members, respectively. The latter were selected from the complete list of all 10554 stars in the field and 
represents all stars that failed at least one of our PMS membership criteria described in the previous sections. From the 
spatial distribution of the non-cluster members one can clearly identify the extent of the dust cloud located towards 
NGC\,2264. When comparing Fig.~\ref{location_pv_iv} with Fig.~\ref{pos_back} it is evident that most PMS stars are located 
in the region with the highest background extinction, i.~e. close to the dense gas and dust out of which they have probably 
been formed.

\subsection{Colour-magnitude diagrams of the periodic and irregular variables}\label{CMDalles}

In Fig.~\ref{FHD} we show the $I_{\mathrm{C}}$ vs ($R_{\mathrm{C}}-I_{\mathrm{C}}$) colour-magnitude diagram of all 
periodic and irregular variables. It is evident that the periodic variable PMS members are much more concentrated around 
a line parallel to the ZAMS than the irregular variable PMS members. The reason for this could be a smaller age range of 
the periodic variables and/or a higher variability of the irregular variables compared to the periodic variables. In 
addition a higher intrinsic extinction of the irregular variables due to circumstellar disks could lead to a larger 
scatter. In the following section the differences between the two subsamples will be discussed in more detail.

In Fig.~\ref{FHD} we also show the $I_{\mathrm{C}}$ vs ($R_{\mathrm{C}}-I_{\mathrm{C}}$) colour-magnitude diagram of all 
(169) previously known PMS stars in NGC\,2264 which we could identify in our sample. It is evident that the new PMS 
variables we found extend to much fainter magnitudes than the previously known PMS stars, i.~e. we found new PMS stars 
mainly in the low mass regime which probably reaches down to the substellar limit of approximately $I_{\mathrm{C}}=
18.5\,\mathrm{mag}$ in NGC\,2264.

\section{The nature of the variability}\label{natvari}

In this section we investigate some general aspects of the stars' variability and investigate which differences in physical 
properties between periodic and irregular variables can be found. 

\subsection{The degree of variability}

We use the standard deviation $\sigma$ of the light curves as an estimation of the degree of variability. The advantage 
of using $\sigma$ (see Sect.~\ref{iv}) rather than the peak-to-peak (ptp) variation of the stars for the characterisation of 
the variability is that $\sigma$ is more robust to outliers in the light curves (e.~g. due to cosmic rays). In order to 
account for those statistical outliers in the light curves we have equated the ptp variation for each star with the magnitude 
difference between the third highest and the third lowest data point. Both quantities $\sigma$ and ptp variation are given in 
Table~\ref{ergebnis} for each star. We find that the ptp variation is typically three times the standard deviation but 
it can vary between two and five times the standard deviation in some extreme cases.
\begin{figure}[t]
  \resizebox{\hsize}{!}{\includegraphics{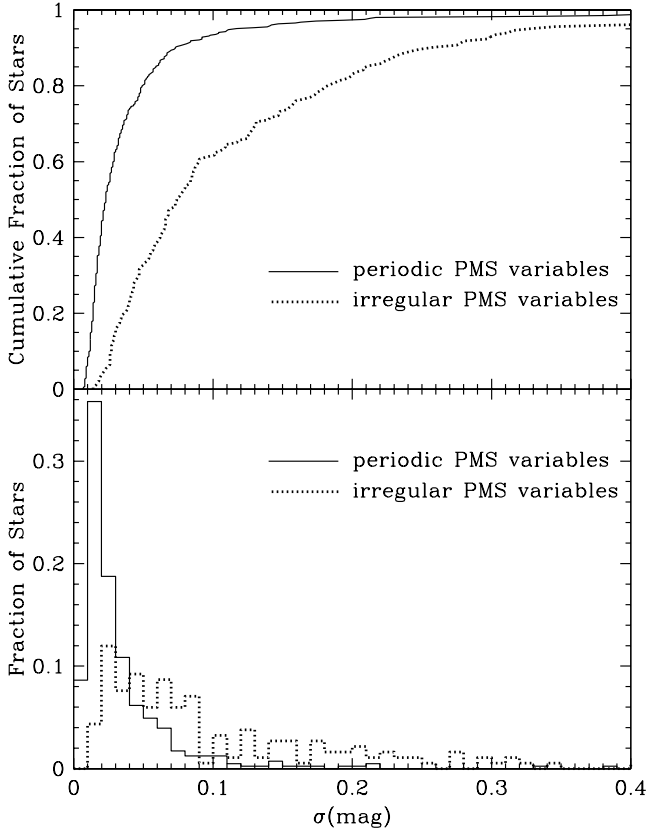}}
       \caption{{\bf (a)} The top panel shows the cumulative distribution of the standard deviation $\sigma$ for 
		all 405 periodic and 184 irregular variable PMS stars respectively.
		{\bf (b)} The lower panel shows the histogram of $\sigma$ for the two different samples of stars in (a).}
    \label{sigma_vert}
\end{figure}

In Fig.~\ref{sigma_vert} we show the cumulative distribution of the standard deviation $\sigma$ 
for all 405 periodic and 184 irregular PMS variables. It is evident from this plot that about 
half of the periodic variables have $\sigma \leq 0.02\,\mathrm{mag}$ while only $5.4\%$ (10/184) 
irregular variables show this small level of variability. On the other hand only $2.7\%$ of 
the periodic variables have a $\sigma \geq 0.2\,\mathrm{mag}$ but $14.7\%$ of the irregular 
variables exceed this level, i.~e. the fraction of stars with a small $\sigma$ is always higher 
for the periodic variables. 

This excess of periodic low amplitude stars is not surprising because the advantage of the 
periodogram analysis is that it is sensitive even for low amplitude variations. This results 
from the fact that the power from photometric errors (white noise) is distributed over the 
whole frequency domain of the power spectrum. Therefore even low amplitude variations close 
to (and below) the photometric errors can be detected. The $\chi ^2$ test on the other hand 
is able to detect variability only if the variations in a light curve are inconsistent with 
the photometric errors. Our cutoff level of a probability $P_{\mathrm{vari}} = 99.9\%$ 
corresponds to deviations of more than $3\sigma$. Therefore we expect to find more periodic 
variables than irregular variables among the low amplitude variables.

In order to take any bias resulting from the different variability analysis methods into 
account we selected only those 264 periodic variables which are also variable according 
to the $\chi^2$ test and compared their distribution with that of the irregular variables. 
We found again that the two distributions are significantly different, as confirmed by a 
Kolmogorov-Smirnov test (Press et al. \cite{press}) which shows that the probability that 
the two distributions are equivalent is less than $5\times 10^{-15}$. A further discussion
on this follows in Sects.~\ref{sec_haindex} and \ref{fractions}.

A similar result was found in the ONC (Herbst et al. \cite{herbst2002}) and in other 
T~Tauri star associations. Herbst et al. (\cite{herbst1994}) have shown that most 
irregular variables are CTTSs and that they show large variations in their light 
curves. Periodic variables on the other hand are mainly WTTSs and have peak-to-peak 
variations in $I_{\mathrm{C}}$ of less than 0.5\,mag. This would correspond to 
standard deviations of $\sigma = 0.1~\mathrm{to}~0.25\,\mathrm{mag}$ for stars 
in our sample. Herbst et al. (\cite{herbst2002}) furthermore concluded that the 
variations in the stellar brightness of the periodic WTTSs are mainly caused 
by cool spots and the observed maximum peak-to-peak variation of 0.5\,mag is 
interpreted as the maximum possible brightness change which cool spots on the 
surface of a K or M star could cause. The periodic variables with peak-to-peak 
variations larger than 0.5\,mag are believed to be due to hot surface spots 
resulting from mass accretion. In our sample $3.5\%$ of the periodic variables have 
peak-to-peak variations $\geq 0.5\,\mathrm{mag}$ ($\sigma \geq 0.16\,\mathrm{mag}$). 
The irregular variability of CTTSs can have various reasons. Often it is attributed 
to variable mass accretion resulting in hot spots which are not stable in brightness, 
size, and location over a few rotation periods. In addition flare-like activities 
can be an additional source of variability. In reality the situation is probably 
more complicated and it could well be that some WTTSs have small hot spots and that 
CTTSs have cool spots in addition to large hot spots. However, for CTTSs it is much 
more difficult and in many cases even impossible to detect periodic brightness 
modulations caused by cold spots because of the overlying ``noise'' from irregular 
variability.

\subsection{The $H\alpha$ emission index of periodic and irregular variables}\label{sec_haindex}

\begin{figure}
  \resizebox{\hsize}{!}{\includegraphics{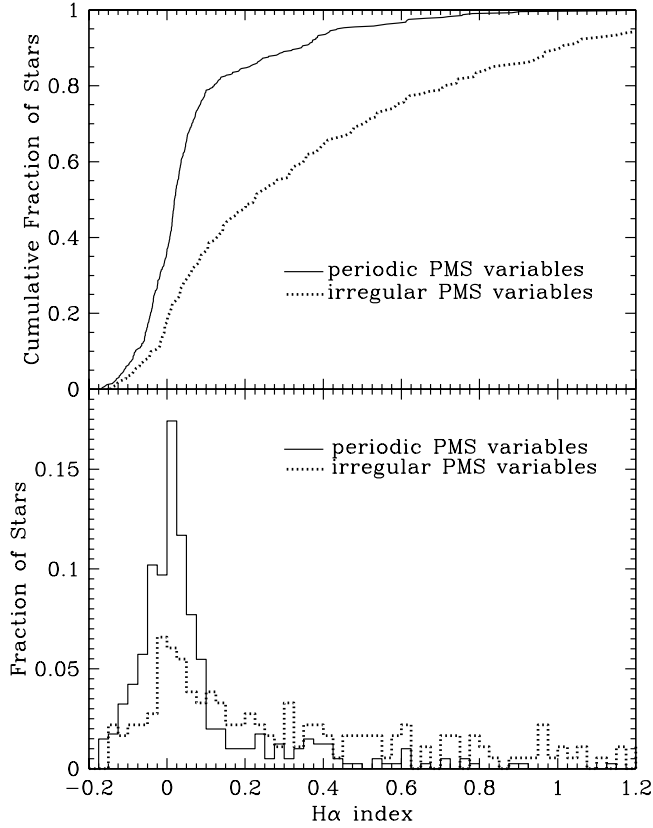}}
       \caption{{\bf (a)} The top panel shows the cumulative frequency distribution 
		of the $H\alpha$ index $\Delta (R_{\mathrm{C}}-H\alpha)$ for all 405 
		periodic and 184 irregular variable PMS stars. {\bf (b)} The lower 
		panel shows the histogram of the $H\alpha$ index for the two different 
		samples of stars in (a).}
    \label{haindex_pv_iv}
\end{figure}
If the fraction of CTTSs is in fact higher among the irregular variables they should also 
show larger $H\alpha$ emission than the periodic variables. In order to investigate the 
$H\alpha$ emission of the stars in the two samples we defined a $H\alpha$ emission index 
$\Delta (R_{\mathrm{C}}-H\alpha)$ of a star by the following equation:	
\begin{equation}		
\Delta (R_{\mathrm{C}}-H\alpha) = (R_{\mathrm{C}}-H\alpha)_{\mathrm{star}} 
				- (R_{\mathrm{C}}-H\alpha)_{\mathrm{locus}},
\end{equation}	 
where the $(R_{\mathrm{C}}-H\alpha)_{\mathrm{locus}}$ is shown as a solid line in 
Fig.~\ref{halphafig} and is given by Eq.~\ref{rhams}. It is the $(R_{\mathrm{C}}-H\alpha)$, 
$(R_{\mathrm{C}}-I_{\mathrm{C}})$ relation of a MS star. The $H\alpha$ index is a 
measurement of the $H\alpha$ emission. Using stars with known $H\alpha$ equivalent 
width ($W_{\lambda}(H\alpha)$) we find (see Paper II) that $83\%$ of the stars with 
an emission index of $\Delta (R_{\mathrm{C}}-H\alpha) \geq 0.1\,\mathrm{mag}$ have 
$W_{\lambda}(H\alpha) \ga 10${\AA} and are therefore likely CTTSs. On the other hand 
about $85\%$ of the stars with  $\Delta (R_{\mathrm{C}}-H\alpha)<0.1\,\mathrm{mag}$ 
have $W_{\lambda}(H\alpha)<10${\AA} and are therefore most likely WTTS. 

In Fig.~\ref{haindex_pv_iv} we show the cumulative distribution and the 
histogram of the $H\alpha$ index for periodic and irregular variables.
It is evident that the $H\alpha$ emission index of the periodic variables is 
concentrated around zero and that only $22\%$ of these stars exceed an $H\alpha$ 
index of 0.1\,mag. On the other hand the $H\alpha$ emission index of the irregular 
variables is distributed over a larger range and for $68\%$ of the irregular 
variables it is above the critical level of 0.1\,mag. 

This supports the interpretation that the fraction of CTTSs among the irregular 
variables is higher than the fraction of CTTSs among the periodic variables and 
vice versa for the WTTSs. In Sect.~\ref{fractions} we will quantify how strongly 
the sample of periodic variables is biased towards the WTTSs by comparing the 
fractions of CTTSs and WTTSs among the periodic variables with the expected 
fractions.

\begin{figure}
  \resizebox{\hsize}{!}{\includegraphics{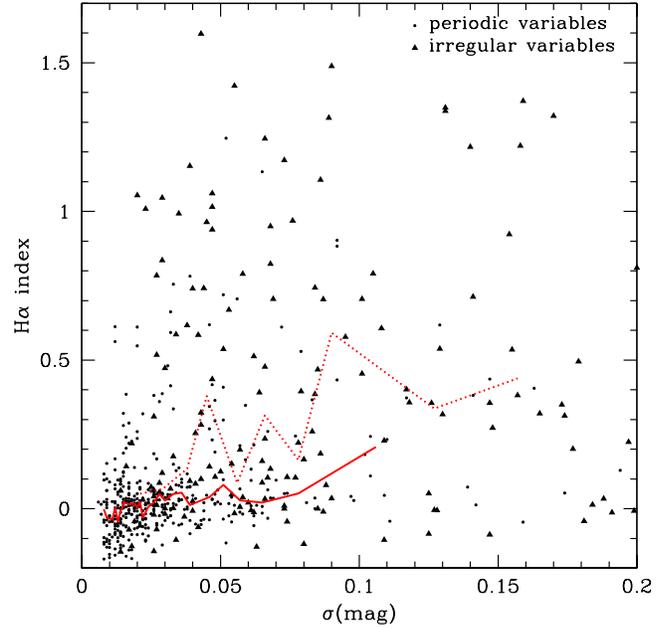}}
       \caption{The $H\alpha$ index as a function of the degree of variability 
		measured by the standard deviation $\sigma$ for all 405 periodic 
		and 184 irregular variable PMS stars. Circles represent periodic 
		variable stars while triangles represents irregular variables. 
		The dashed and solid lines represent the medians of the $H\alpha$ 
		index for the irregular and periodic variables respectively in 
		$\sigma$-bins of variable width with each containing 14 data points.}
    \label{sigma_haindex}
\end{figure}

\subsection{The correlation between $\sigma$ and the $H\alpha$ index}\label{corosigha}

In order to investigate this interpretation further we look for a correlation 
between the degree of variability of the stars and their $H\alpha$ index which 
is a measurement of their accretion and/or chromospheric activity. In 
Fig.~\ref{sigma_haindex} we show the $H\alpha$ index and its median as a 
function of $\sigma$ for periodic and irregular variables respectively. It shows 
that for both samples there is an increase of the median $H\alpha$ index with 
increasing $\sigma$. Each median was calculated in $\sigma$ bins of variable 
width containing 14 data points. Only the range $\sigma \leq 0.22\,\mathrm{mag}$ 
is considered here since there are insufficient data points for larger values of 
$\sigma$. A Spearman rank-order correlation test (Press et al. \cite{press}) 
indicates that the probability that $\sigma$ and the $H\alpha$ index are not 
correlated is less than $6\times 10^{-28}$ if we use the joined sample of 
periodic and irregular variable stars. However, this high probability that there 
is a correlation between the two quantities is dominated by the periodic sample. 
If we use only the periodic sample the probability that $\sigma$ and the $H\alpha$
index are not correlated is less than $9\times 10^{-15}$. On the other hand the 
probability that $\sigma$ and the $H\alpha$ index are correlated for the irregular 
variables is only $0.863$. 

It is interesting that variability in the periodic variables is strongly correlated 
with the $H\alpha$ index while there is no evidence for a correlation for the 
irregular variables. One possible reason is that the sizes and numbers of cool spots 
(i.~e. the total spot area) on periodic variables (WTTSs) are correlated with the 
chromospheric activity as in other active late type stars, i.~e. periodic variables 
with more spots have more active regions on their surface and therefore stronger 
chromospheric $H\alpha$ emission. For the irregular variables we see that there is
a large scatter in the $H\alpha$, $\sigma$ relation, and even low amplitude variables 
have large $H\alpha$ indices. This could imply that there are many CTTSs which have 
no large variations in their mass accretion rates. In addition these stars must have 
hot spot patterns which do not cause any large light modulations (e.~g. accretion rings
symmetric to the rotation axis, see e.~g. Mahdavi \& Kenyon \cite{mahdavi}).

\subsection{The $(V-R_{\mathrm{C}})$ vs $(R_{\mathrm{C}}-I_{\mathrm{C}})$ colour-colour-diagram}

\begin{figure}
  \resizebox{\hsize}{!}{\includegraphics{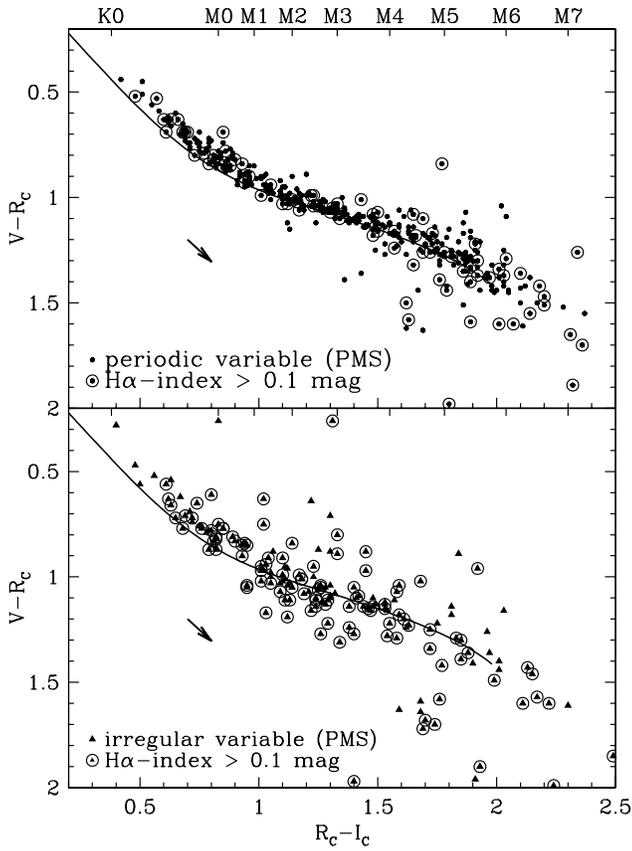}}
       \caption{The $(V - R_{\mathrm{C}})$ vs $(R_{\mathrm{C}} - I_{\mathrm{C}})$ 
		colour-colour-diagram for periodic (top	panel) and irregular variables 
		(bottom panel) that passed both of our two PMS tests. Those stars which 
		have an $H\alpha$-index$\geq 0.1$ and therefore show the properties of 
		accreting CTTSs are marked with a surrounding circle. The solid lines in 
		both panels represents the ZAMS. The arrow length and direction indicates 
		the mean reddening towards NGC\,2264.
		}
    \label{vrir}
\end{figure}

In Fig.~\ref{vrir} we show the $(V - R_{\mathrm{C}})$ vs $(R_{\mathrm{C}}-I_{\mathrm{C}})$ 
colour-colour-diagram for periodic and irregular variables which passed both of our two 
PMS tests. The stars with an $H\alpha$-index of $\Delta (R_{\mathrm{C}}-H\alpha)\leq 0.1$ 
are marked separately. It is evident that the scatter is much larger among the irregular 
variables compared to the periodic variables. Agreement with the ZAMS is good, although
deviations due to the smaller $\log\,g$ values of the PMS stars are expected. The data 
points below the MS (i.~e. with larger $(V - R_{\mathrm{C}})$ colours than a MS star) can 
be explained in both plots by embedded stars which are highly reddened with up to eight 
times the mean reddening. This interpretation is supported by the analysis of Rebull et al. 
(\cite{rebull2002}). They have determined an average $E(R_{\mathrm{C}}-I_{\mathrm{C}})=
0.1\pm 0.02\,\mathrm{mag}$ in NGC\,2264, but showed that some stars in NGC\,2264 are 
reddened by up to $E(R_{\mathrm{C}}-I_{\mathrm{C}})=1\,\mathrm{mag}$.

\section{``Non-variable'' PMS members with strong $H\alpha$ emission}\label{PMScand}

\begin{figure}
  \resizebox{\hsize}{!}{\includegraphics{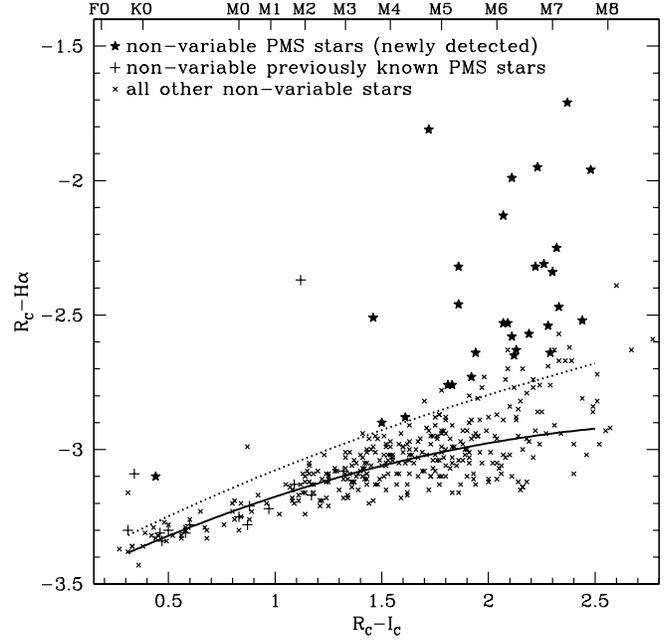}}
       \caption{The $(R_{\mathrm{C}}-H\alpha)$ vs $(R_{\mathrm{C}} - I_{\mathrm{C}})$ 
		colour-colour-diagram for all ``non-variable'' stars which passed the first
		PMS test (see Sect.~\ref{member}) including previously known ``non-variable'' 
		PMS members. The dotted line represents the modified second selection 
		criterion for ``non-variable'' PMS stars (see text). All ``non-variable'' stars 
		which are located by more than their error in $(R_{\mathrm{C}}-H\alpha)$ 
		above this level are classified as PMS candidates.
		}
    \label{selcand}
\end{figure}
Based on our periodic and irregular variability study we have selected a total number 
of 584 PMS members (405 periodic and 184 irregular variables) using the two selection 
criteria described in Sect.~\ref{member}. However, there may exist low-amplitude PMS 
stars, for which we could detect no significant variability. Reasons for a small degree 
of variability in WTTSs could be a more or less equally distributed pattern of many spots 
causing only a small modulation of the stellar brightness and in CTTSs there could be 
phases of relatively stable mass accretion. To simplify matters we call these stars 
``non-variable'' PMS stars. In order to search for ``non-variable'' PMS stars both PMS 
tests described in Sect.~\ref{member} were applied to all stars which were not detected 
to be variable. ``Non-variable'' PMS stars were selected if they passed both tests and 
have in addition strong $H\alpha$ emission.

In Fig.~\ref{selcand} we show the $(R_{\mathrm{C}}-H\alpha)$ vs 
$(R_{\mathrm{C}}-I_{\mathrm{C}})$ colour-colour-diagram for all of our non-variable 
stars which passed the first selection criterion. From these stars we selected only 
stars which showed an enhanced $H\alpha$ emission using a more conservative threshold 
in the colour-colour diagram. This threshold is indicated by the dotted line in 
Fig.~\ref{selcand} which represents the upper envelope of the MS locus (solid line). 
The dotted line is given by 
\begin{equation}\label{upperlevel}
(R_\mathrm{C}-H\alpha)_{\mathrm{high}} = (R_\mathrm{C}-H\alpha)_{\mathrm{locus}}
				       + 1.65\times \delta,
\end{equation} 
(for the definition of $\delta$ and $(R_\mathrm{C}-H\alpha)_{\mathrm{locus}}$ see 
Sect.~\ref{selCCD}). This upper discrimination level is equal to the lower discrimination 
level used in Sect.~\ref{selCCD} mirrored at the PMS locus.

We consider only stars as "non-variable" PMS members if they are located more than 
their photometric error $\Delta_{(R_{\mathrm{C}}-H\alpha)}$ above the upper 
discrimination line, i.~e. $(R_\mathrm{C}-H\alpha)-\Delta_{(R_\mathrm{C}-H\alpha)}
\geq (R_\mathrm{C}-H\alpha)_{\mathrm{high}}$. In total 32 stars of the non-variables 
with $(R_{\mathrm{C}}-I_{\mathrm{C}}) \leq 2.5\,\mathrm{mag}$ in Fig.~\ref{selcand} 
passed this additional PMS test. Out of these "non-variable" PMS members 2 stars 
(no. 5941 and no. 6164) were classified as PMS stars prior to this study. Another 
12 previously known ``non-variable'' PMS members did not pass this additional test. 
The 30 newly found stars are separately marked in Fig.~\ref{selcand} and listed in 
Table~\ref{pmscandtab}. 

As one can see from Fig.~\ref{selcand} most ($83\%$) of the new "non-variable" PMS 
members have $(R_{\mathrm{C}}-I_{\mathrm{C}})\le 1.8\,\mathrm{mag}$. This can be 
understood if we consider that the photometric errors for these stars are typically 
larger compared to bluer (i.~e. brighter) stars. Therefore any variations in the 
light curve are harder to detect.
\begin{table*}[t]
 \begin{minipage}[b][][t]{235mm}
  \centering
  \caption{New "non-variable" PMS stars in NGC\,2264 selected using the PMS-test I and a slightly modified PMS-test II (see text).
	   $P_{vari}$ is the probability that the star is variable according to our $\chi^2$ test. The other columns are the 
	   same as in Table \ref{ergebnis}.} 
  \begin{center}
  \label{pmscandtab}
  \begin{tabular*}{235mm}[b]{@{\extracolsep\fill}c@{\extracolsep\fill} c@{\extracolsep\fill} c@{\extracolsep\fill} c@{\extracolsep\fill}c @{\extracolsep\fill}c@{\extracolsep\fill}c @{\extracolsep\fill}c@{\extracolsep\fill}c @{\extracolsep\fill}c@{\extracolsep\fill}c @{\extracolsep\fill}c @{\extracolsep\fill}c @{\extracolsep\fill}r @{\extracolsep\fill}l} 
	\hline
	\hline
 Star & $\alpha (J2000)$&$\delta (J2000)$ & {$I_{\mathrm{C}}$}& {err} &{ $V-I_{\mathrm{C}}$} & {err} & {$R_{\mathrm{C}}-I_{\mathrm{C}}$} & {err} & {$R_{\mathrm{C}}-H\alpha$} & {err} & {$\sigma$} & {ptp}  & $ P_{vari} $ & {cross id} \\ 
	\hline
 3130 & 6:40:21.32 & 9:48:26.5 & 19.00 & 0.03 &\dotfill&\dotfill& 2.26 & 0.03 &-2.31 & 0.10 & 0.040 & 0.17 &  0.0015 &\\
 3455 & 6:40:26.95 & 9:49:55.7 & 18.28 & 0.02 &\dotfill&\dotfill& 2.19 & 0.02 &-2.57 & 0.05 & 0.014 & 0.05 & 32.8153 & R2324   \\
 3646 & 6:40:29.41 & 9:47:37.5 & 16.83 & 0.02 & 3.26   &  0.06  & 2.07 & 0.02 &-2.13 & 0.07 & 0.017 & 0.05 & 99.3415 & R2401   \\
 3648 & 6:40:29.42 & 9:41:04.4 & 18.60 & 0.01 & 3.39   &  0.07  & 2.07 & 0.03 &-2.53 & 0.06 & 0.017 & 0.06 &  0.0005 &\\
 3680 & 6:40:29.89 & 9:50:20.2 & 16.79 & 0.01 & 2.62   &  0.01  & 1.50 & 0.01 &-2.90 & 0.01 & 0.007 & 0.03 &  0.0000 &\\
 3840 & 6:40:31.92 & 9:50:16.2 & 19.42 & 0.03 &\dotfill&\dotfill& 2.28 & 0.05 &-2.54 & 0.15 & 0.034 & 0.12 &  1.6468 &\\
 4323 & 6:40:38.30 & 9:29:25.8 & 17.71 & 0.04 & 3.06   &  0.02  & 1.81 & 0.01 &-2.76 & 0.05 & 0.157 & 0.65 &  0.0363 &\\
 4877 & 6:40:45.89 & 9:39:19.6 & 19.19 & 0.04 & 2.07   &  0.08  & 2.09 & 0.09 &-2.53 & 0.18 & 0.035 & 0.13 &  0.0031 &\\
 4924 & 6:40:46.68 & 9:48:13.0 & 16.92 & 0.01 & 3.05   &  0.02  & 1.86 & 0.02 &-2.46 & 0.02 & 0.007 & 0.03 &  0.0000 & R2959   \\
 4960 & 6:40:47.17 & 9:28:50.4 & 11.45 & 0.01 & 0.88   &  0.01  & 0.44 & 0.01 &-3.10 & 0.06 & 0.014 & 0.06 &  0.0000 & Y2525,R2979,P67,MX,114\\
 5456 & 6:40:54.74 & 9:54:57.7 & 19.23 & 0.02 &\dotfill&\dotfill& 2.23 & 0.07 &-1.95 & 0.44 & 0.034 & 0.17 &  0.7261 &\\
 5567 & 6:40:56.64 & 9:32:20.2 & 19.33 & 0.03 &\dotfill&\dotfill& 2.30 & 0.05 &-2.34 & 0.16 & 0.024 & 0.10 &  0.0000 &\\
 5570 & 6:40:56.73 & 9:38:10.7 & 18.98 & 0.02 &\dotfill&\dotfill& 2.48 & 0.07 &-1.96 & 0.10 & 0.031 & 0.14 & 59.4923 &\\
 5627 & 6:40:59.39 & 9:31:00.2 & 17.60 & 0.02 & 3.14   &  0.09  & 1.86 & 0.02 &-2.32 & 0.03 & 0.022 & 0.07 & 99.3212 &\\
 5631 & 6:40:59.53 & 9:34:44.2 & 19.23 & 0.02 &\dotfill&\dotfill& 2.22 & 0.05 &-2.32 & 0.09 & 0.044 & 0.16 &  3.4192 &\\
 5982 & 6:41:07.56 & 9:41:34.1 & 16.07 & 0.01 & 2.53   &  0.04  & 1.46 & 0.01 &-2.51 & 0.10 & 0.014 & 0.05 &  2.5542 &R3568   \\
 6047 & 6:41:09.23 & 9:29:05.0 & 18.48 & 0.02 &\dotfill&\dotfill& 2.11 & 0.03 &-1.99 & 0.06 & 0.019 & 0.07 &  0.2423 &\\
 6169 & 6:41:13.11 & 9:24:36.9 & 16.44 & 0.01 & 2.79   &  0.01  & 1.61 & 0.01 &-2.88 & 0.01 & 0.006 & 0.03 &  0.0000 &Y3810,R3716,MX264\\
 6190 & 6:41:13.77 & 9:29:32.5 & 17.64 & 0.02 & 3.53   &  0.04  & 2.12 & 0.02 &-2.65 & 0.03 & 0.009 & 0.04 &  0.0394 & R3729   \\
 6306 & 6:41:17.38 & 9:35:56.5 & 19.28 & 0.03 &\dotfill&\dotfill& 2.44 & 0.05 &-2.52 & 0.12 & 0.030 & 0.13 &  0.0551 &\\
 6354 & 6:41:18.66 & 9:49:02.7 & 18.20 & 0.01 & 3.31   &  0.04  & 1.92 & 0.01 &-2.73 & 0.06 & 0.017 & 0.06 &  0.4781 &\\
 6355 & 6:41:18.69 & 9:32:52.7 & 19.03 & 0.02 &\dotfill&\dotfill& 2.32 & 0.07 &-2.25 & 0.08 & 0.026 & 0.12 &  0.0000 &\\
 6359 & 6:41:18.77 & 9:35:19.6 & 19.45 & 0.04 &\dotfill&\dotfill& 2.33 & 0.07 &-2.47 & 0.16 & 0.033 & 0.14 &  0.3219 &\\
 6490 & 6:41:21.44 & 9:53:01.9 & 17.86 & 0.01 & 2.94   &  0.03  & 1.72 & 0.03 &-1.81 & 0.04 & 0.018 & 0.07 & 43.0053 &\\
 6607 & 6:41:23.29 & 9:32:30.7 & 18.70 & 0.03 & 3.33   &  0.07  & 2.13 & 0.02 &-2.63 & 0.06 & 0.025 & 0.10 &  0.1210 &\\
 6668 & 6:41:24.02 & 9:26:53.0 & 17.61 & 0.01 & 3.08   &  0.03  & 1.83 & 0.01 &-2.76 & 0.03 & 0.008 & 0.03 &  0.0082 &\\
 6805 & 6:41:25.93 & 9:30:26.6 & 17.89 & 0.01 & 3.45   &  0.04  & 2.11 & 0.03 &-2.58 & 0.05 & 0.014 & 0.05 &  2.3883 & R4001   \\
 7037 & 6:41:28.61 & 9:53:58.5 & 17.14 & 0.01 & 3.23   &  0.02  & 1.94 & 0.02 &-2.64 & 0.07 & 0.013 & 0.05 & 12.7790 &\\
 7075 & 6:41:29.16 & 9:28:48.6 & 19.39 & 0.03 &\dotfill&\dotfill& 2.37 & 0.08 &-1.71 & 0.10 & 0.029 & 0.11 &  0.0000 &\\
10865 & 6:42:03.92 & 9:49:02.2 & 18.94 & 0.02 & 3.49   &  0.13  & 2.29 & 0.03 &-2.64 & 0.08 & 0.037 & 0.14 & 45.2014 &\\
  \hline
  \end{tabular*}
  \end{center}
  \end{minipage}
\end{table*}

In total we have identified 621 PMS stars in NGC\,2264 including 405 periodic 
variable 184 irregular variable and 32 "non-variable" stars with relatively 
strong $H\alpha$ emission. Compared to the 182 previously known PMS stars and 
PMS candidates reported by Park et al. \cite{park} (not including 26 previously 
known massive OB stars) we have increased the number of known PMS stars in 
NGC\,2264 by a factor of 3.4. As already mentioned in Sect.~\ref{pvdatabase} 
most of the newly found PMS stars are low mass stars ($M\la 0.25\,M_{\sun}$) 
with masses probably extending into the substellar regime.

\section{Fraction of variable PMS stars and completeness level of our PMS sample}\label{comleteness}

In this section we estimate the fraction of variable PMS stars in the cluster. 
From this fraction we estimate the completeness level of our sample of variable 
PMS stars in NGC\,2264. Therefore the central question in this section is: what
fraction of PMS stars did we find through our photometric monitoring program 
and the adopted methods in selecting PMS stars?

The fraction of variable cluster members is given by $f_{\mathrm{vari}}=
N_{\mathrm{vari}}/(N_{\mathrm{vari}}+N_{\mathrm{non-vari}})$, where 
$N_{\mathrm{vari}}$ is the number of variable (both periodic and irregular) 
and $N_{\mathrm{non-vari}}$ is the number of "non-variable" PMS stars 
in NGC\,2264. These numbers (in particular $N_{\mathrm{non-vari}}$) are 
well known for CTTSs since the PMS nature of these stars can be confirmed 
by an enhanced $H\alpha$ emission. In addition we expect that there are 
also WTTSs (with no or weak $H\alpha$ emission) which are "non-variable"; 
i.~e. they have light modulations below our detection limit. Using the 
$I_{\mathrm{C}}$ vs $(R_{\mathrm{C}}-I_{\mathrm{C}})$ colour-magnitude 
or $(R_{\mathrm{C}}-H\alpha)$ vs $(R_{\mathrm{C}}-I_{\mathrm{C}})$ 
colour-colour diagrams these stars are indistinguishable from MS
foreground stars. In order to estimate the number of "non-variable" 
WTTSs we therefore have to keep the relative contamination by 
foreground MS stars as low as possible. This is achieved by using only 
the two regions NGC\,2264~N and S (see Sect.~\ref{selCCD}) for our 
analysis rather than the whole observed field. In order to be not 
affected in the following analysis by photometric errors we restrict 
ourselves to stars with $I_{\mathrm{C}} \leq 18.0\,\mathrm{mag}$ and 
$(R_{\mathrm{C}}-I_{\mathrm{C}})\leq 1.8\,\mathrm{mag}$.

In Table~\ref{anteil} we list the number of variable and "non-variable" stars 
in each of the two regions which passed both PMS tests (see Sect.~\ref{member}) 
and are in these colour and magnitude ranges. From the last column of this 
table it is evident that in the two regions NGC\,2264~N and S about $74\%$ of 
the PMS candidates (i.~e. stars that passed the two tests) are variable. About 
half of the PMS candidates show periodic light modulations while only one 
fourth of the PMS candidates are irregular variables. We note that these fractions 
are only lower limits since we assume that all stars in the two regions which 
passed both tests are indeed PMS stars. In particular we assume that the 
contamination by non-members in the variable sample is negligible; i.~e. all 
variables that passed both PMS test are actually PMS stars.

However, if we further assume that the two regions are representative for the 
whole cluster we can conclude that at least $f_{\mathrm{vari}}=74\%$ of the 
PMS stars have been detected to be variable. This lower limit of the fraction 
of detected variable stars is not significantly higher if we consider only 
brighter stars; e.~g. for stars with $I_{\mathrm{C}}\leq 16.0\,\mathrm{mag}$ 
in NGC\,2264~N and S we obtain a fraction of $f_{\mathrm{vari}}=76\%$. 
\begin{table}
\centering
       \caption{The estimated fraction of variable stars in NGC\,2264~N 
		and S with $I_{\mathrm{C}} \leq 18.0\,\mathrm{mag}$ and 
		$(R_{\mathrm{C}}-I_{\mathrm{C}})\leq 2.0\,\mathrm{mag}$. 
		Listed are number and fractions of periodic PMS variables 
		($N_{\mathrm{p-vari}}$), irregular PMS variables 
		($N_{\mathrm{i-vari}}$), "non-variable" PMS candidates 
	 	($N_{\mathrm{non-vari}}$), the total number of all PMS 
		candidates ($N_{\mathrm{PMS-total}}=N_{\mathrm{p-vari}}
		+N_{\mathrm{i-vari}}+N_{\mathrm{non-vari}}$), and the 
		number and fractions of PMS variables ($N_{\mathrm{p-vari}} 
	 	+N_{\mathrm{i-vari}}$) in the two concentrations of PMS 
		stars NGC\,2264\,N and S (for a definition of the PMS 
		candidates see text). The fractions were calculated 
		relative to the total number of PMS star candidates
		($N_{\mathrm{PMS-total}}$) in the two regions.
	}
  	\label{anteil}
  	\begin{tabular}{lrrr} 
	\hline
	\hline
	sample of stars                   & \multicolumn{1}{@{\extracolsep\fill}c}{North}& 
					    \multicolumn{1}{@{\extracolsep\fill}c}{South}& 
					    \multicolumn{1}{@{\extracolsep\fill}c}{North \& South} \\
	\hline
	$N_{\mathrm{PMS-total}}$ \dotfill & \multicolumn{1}{@{\extracolsep\fill}c}{110 \dotfill} &
					    \multicolumn{1}{@{\extracolsep\fill}c}{120 \dotfill} &
					    \multicolumn{1}{@{\extracolsep\fill}c}{230 \dotfill}  \\	
	$N_{\mathrm{non-vari}}$  \dotfill & 25 ($22.7\%$) & 36 ($30.0\%$) &  61 ($26.5\%$)\\	
	$N_{\mathrm{p-vari}}$    \dotfill & 56 ($50.9\%$) & 54 ($45.0\%$) & 110 ($47.8\%$)\\	
	$N_{\mathrm{i-vari}}$    \dotfill & 29 ($26.4\%$) & 30 ($25.0\%$) &  59 ($25.7\%$)\\	
	\hline
	variables                \dotfill & 85 ($77.3\%$) & 84 ($70.0\%$) & 191 ($73.5\%$)\\	
	\hline 
  	\end{tabular}
\end{table}

Since we expect {\it all} PMS stars in the cluster to be variable at some 
level our variability study is complete at a $74\%$ level for stars with 
$I_{\mathrm{C}}\leq 18.0\,\mathrm{mag}$ and $(R_{\mathrm{C}}-I_{\mathrm{C}}) 
\leq 1.8\,\mathrm{mag}$. As we will show in the following section the 
completeness level differs for WTTSs and CTTSs. While nearly all ($95\%$) of 
the CTTSs are found to be variable only $68\%$ of the WTTSs are found to be 
variable.

In summary we conclude that our method of photometric monitoring is a very 
powerful tool for finding most PMS stars in our cluster since at least three 
quarters of the whole PMS population in NGC\,2264 (with $(R_{\mathrm{C}}-
I_{\mathrm{C}}) \leq 1.8\,\mathrm{mag}$) could be identified this way. Therefore,
our database of periodic and irregular variables should be a representative 
subset of the PMS stars in NGC\,2264.

\section{Fractions of periodic variables among CTTSs and WTTSs}\label{fractions}

In the following we try to estimate the fraction of periodic variables among the 
WTTSs and CTTSs in order to investigate how strongly the period distribution of 
NGC\,2264 is biased by the non-detection of periods in these two groups of stars. 
Again we restrict our analysis to stars with $I_{\mathrm{C}}\leq 18.0\,\mathrm{mag}$ 
and $(R_{\mathrm{C}}-I_{\mathrm{C}}) \leq 1.8\,\mathrm{mag}$. Furthermore we consider
in this section only stars that passed both PMS tests.

As already outlined in Sect.~\ref{natvari} the periodic variables are biased towards 
the WTTSs since it is much harder to detect periodic brightness modulations of 
CTTSs in the presence of a superimposed irregular variability. However, we will show 
below that this bias towards the WTTSs will be partially compensated by those WTTSs 
which have brightness modulations below our detection limit and have therefore not 
been found.

As in Sects.~\ref{natvari} and \ref{PMScand} we regard all stars with enhanced 
$H\alpha$ emission measured by the $H\alpha$ index as CTTSs. This is the case for 
$N_{\mathrm{CTTS}}=145$ stars in our whole investigated region. While 89 ($61.4\%$) 
of these stars are irregularly variable only 49 ($33.8\%$) are periodically variable
and 7 ($4.8\%$) are ``non-variable''. Thus only for one third (49/145) of the CTTSs 
periods are detectable. 

It is more difficult to estimate the total number of WTTSs in the cluster 
($N_{\mathrm{WTTS}}$) since ``non-variables'' WTTSs are indistinguishable from 
MS foreground stars in the $I_{\mathrm{C}}$ vs $(R_{\mathrm{C}}-I_{\mathrm{C}})$ 
colour-magnitude and in the $(R_{\mathrm{C}}-H\alpha)$ vs 
$(R_{\mathrm{C}}-I_{\mathrm{C}})$ colour-colour diagram. However, with the 
results of the previous section we can relatively easily estimate the total 
number of PMS stars in the cluster ($N_{\mathrm{PMS}}$) and since 
$N_{\mathrm{PMS}}=N_{\mathrm{CTTS}}+N_{\mathrm{WTTS}}$ we can determine 
$N_{\mathrm{WTTS}}$ indirectly.

Therefore we first estimate $N_{\mathrm{PMS}}$. In total 465 stars with 
$(R_{\mathrm{C}}-I_{\mathrm{C}}) \leq 1.8\,\mathrm{mag}$ and $I_{\mathrm{C}} \leq 
18.0\,\mathrm{mag}$ are periodic or irregular variables. Since at least $73.5\%$ 
of the PMS stars are variable (see Sect.~\ref{comleteness}) we get $N_{\mathrm{PMS}} 
=465/0.735 \simeq 630$. With $N_{\mathrm{WTTS}}=N_{\mathrm{PMS}}-N_{\mathrm{CTTS}}$ 
we get $N_{\mathrm{WTTS}}=630-145=485$ which is an upper limit since 0.735 is a lower
limit.

Of the periodic variables with $(R_{\mathrm{C}}-I_{\mathrm{C}})\leq 1.8\,\mathrm{mag}$ 
and $I_{\mathrm{C}} \leq 18.0\,\mathrm{mag}$ 265 have a small $H\alpha$ index and are 
therefore classified as WTTSs. In addition 62 irregular variables are classified as 
WTTSs. Using these numbers we conclude that for at least $55\%$ (265/485) of the WTTSs 
in NGC\,2264 we have detected a period and at least $13\%$ (62/485) of the WTTSs are 
irregular variable; i.~e. for $68\%$ of the WTTSs we detected variability.

How do the measured period fractions of $55\%$ and $34\%$ for the WTTSs and CTTSs 
respectively influence the final period distribution? To answer this question let 
us first assume that the periods for {\it all} (estimated) 630 PMS stars in the 
cluster with $(R_{\mathrm{C}}-I_{\mathrm{C}})\leq 1.8\,\mathrm{mag}$  and 
$I_{\mathrm{C}} \leq 18.0\,\mathrm{mag}$ could be measured. Using $N_{\mathrm{CTTS}}=145$
and $N_{\mathrm{WTTS}}=485$ (see above) we get that $23\%$ (145/630) of the stars 
are CTTSs while the fraction of WTTSs is $77\%$ (485/630). In our study $16\%$ 
(49/341) of the periodic variables in these colour and magnitude ranges are CTTSs 
and $84\%$ (265/314) of the periodic variables are WTTSs. From these numbers we 
conclude that the final period distribution is slightly biased towards WTTSs.

\section{Summary \& Conclusions}\label{summary}

We have carried out an extensive search for rotation periods and variability in 
NGC\,2264. The main results of this investigation are as follows:

\begin{enumerate}

\item
We have obtained absolute photometry in the Cousins $V$, $R_{\mathrm{C}}$, and 
$I_{\mathrm{C}}$ bands and instrumental $H\alpha$ magnitudes for about 10600 stars 
with $9.8\,\mathrm{mag} \leq I_{\mathrm{C}}\leq 21\,\mathrm{mag}$ in NGC\,2264. 
Relative light curves with more than 80 data points each were obtained for each 
of these stars in the $I_{\mathrm{C}}$ band. Two different periodogram analysis 
techniques and a $\chi^2$ test yielded a sample of 543 periodic and 484 irregular 
variables with $11.4\,\mathrm{mag} \leq I_{\mathrm{C}} \leq 19.7\,\mathrm{mag}$ 
in the observed field.

\item
In order to check the PMS nature of the periodic and irregular variables two 
different PMS selection criteria were applied. These criteria are their 
locations in the $I_{\mathrm{C}}$ vs $(R_{\mathrm{C}}-I_{\mathrm{C}})$ 
colour-magnitude diagram and their location in the $(R_{\mathrm{C}}-H\alpha)$ 
vs $(R_{\mathrm{C}}-I_{\mathrm{C}})$ colour-colour diagram. In this way 405 
periodic and 184 irregular variable PMS stars could be selected from the 
sample. This is a enormous increase in the number of known rotation periods 
compared to the about 30 published periods prior to this study. The rotation 
periods of the periodic variables are typically between a quarter of a day 
and 15 days.

\item
Using an enhanced $H\alpha$ emission index as a PMS indicator we could identify 
an additional 35 PMS stars which are "non-variable" according to our study. Together 
with the 589 variable PMS stars we found a total of 624 PMS stars in NGC\,2264. 
Before our study only 182 PMS members and candidates were known. Most of our newly 
found PMS stars are fainter than $I_{\mathrm{C}}\simeq 15\,\mathrm{mag}$ and of 
late spectral type ($\ga$M2).

\item
Comparison of the distribution of the standard deviation $\sigma$ in the light 
curves as well as the distribution of the $H\alpha$ emission index for periodic 
variables with the corresponding distributions for the irregular variables strongly 
suggests that our method for determining the rotation periods preferentially 
selects weak-line T~Tauri stars (WTTSs) while classical T~Tauri stars (CTTSs) are 
underrepresented. This interpretation is confirmed by the calculated fractions of 
periodic variables among WTTSs and CTTSs which, however,  show that the bias of 
the measured rotation periods towards WTTSs is relatively small. We have estimated 
that $23\%$ of the PMS stars in the cluster are CTTSs but only $15\%$ of the periodic 
variables are CTTSs. The corresponding numbers for WTTS are $77\%$ and $85\%$ 
respectively.

\item
The different properties of the two subclasses of T Tauri stars (CTTSs and WTTSs) 
are supposed to be due to the different mechanisms which produce the variability. 
While the variability of WTTSs is likely due to large (magnetically) cool spots 
the variability of CTTSs is believed to be mainly caused by hot spots resulting 
from accretion flows onto the star. On CTTSs the periodic brightness variations 
resulting from cool spots (which are believed to exist on their surface too) are 
often undetectable due to the ``noise'' from the irregular variability. Determining 
the rotational periods for a larger fraction of CTTSs by photometric monitoring 
would therefore require more extensive observations (e.~g. larger time base and 
dense sampling). 

\item
We have estimated that we have detected variability among at least $70\%$ 
of the PMS stars (defined according to our PMS tests) in NGC\,2264 with 
$I_{\mathrm{C}}\leq 18.0\,\mathrm{mag}$ and 
$(R_{\mathrm{C}}-I_{\mathrm{C}})\leq 1.8\,\mathrm{mag}$. From these numbers 
it is evident that extensive photometric monitoring with a similar accuracy 
as in our study is a powerful tool for finding most PMS stars. We emphasise 
that this method is highly efficient only if there is enough data with a 
proper sampling to allow a search for periodic variables because many PMS 
members were identified by their periodicity and not just by their variability.
We conclude that the sample of variable PMS stars obtained in this study is a 
representative subset of the cluster members and it is not expected that 
additional monitoring programs will substantial increase the number of 
known PMS stars at least for stars with $I_{\mathrm{C}}\leq 18.0\,\mathrm{mag}$.

\end{enumerate}

\begin{acknowledgements}
We thank Luisa Rebull, Russ Makidon, and Steven Strom for providing their partly unpublished photometric data and their 
unpublished rotational data. We also thank Eric Young for providing unpublished spectral types. The authors thank Jochen
Eisl\"offel for comments on the manuscript. 
\end{acknowledgements}

\appendix 
\section{The CLEAN algorithm}\label{appcl}
The CLEAN algorithm we used in Section \ref{clper} subtracts the window function iteratively from the observed, or raw, 
spectrum in the following way. First the spectral window function 
\begin{equation}
W(\nu)=\frac{1}{N}\sum_{n=1}^N e^{-2\pi i \nu t_n}
\end{equation}
and the observed spectrum 
\begin{equation}
D(\nu)=\frac{1}{N}\sum_{n=1}^N d(t_n) e^{-2\pi i \nu t_n}
\end{equation}
are calculated at discrete frequencies with a spacing of $\Delta\nu= 1/(10\,T)$, where $T=t_1 - t_N$ is the total length
of the data span. In our case $T$ is about 62.8 days and the resolution of our CLEAN periodograms is therefore 
$\Delta\nu= 0.0016/d$.

The input into the {\it i}th iteration is the spectrum $D_i$, where the iteration process starts with the (complex) raw
spectrum $D_{i=1}\equiv D$. During each iteration step $i=1,2,\ldots,N$ the algorithm determines the frequency $\nu_i$
and the power $|D_i(\nu_i)|$ of the highest peak in the spectrum $D_i$. Assuming that this peak represents a true period
in the signal the aliasing powers $A_i$ at other frequencies are calculated with
\begin{equation}\label{aieq}
A_i(\nu)=c_iW(\nu-\nu_i) + (c_i)^\ast W(\nu+\nu_i)
\end{equation}
where $c_i$ is the $i$th complex {\it clean component}, that represents the amplitude of the peak and is defined as
\begin{equation}\label{gdef}
c_i = g\frac{D_i(\nu_i) - D_i^{\ast}(\nu_i) W(2\nu_i)}{1-|W(2\nu_i)|^2}.
\end{equation}
(see Eq. (23) and (25) in Roberts et al. \cite{roberts}). The $c_i$ were used in Eq.~\ref{aieq} for the calculation of 
the aliasing powers rather than the power $|D_i(\nu_i)|$ of the peak since the peak itself is affected by aliasing of the 
corresponding negative peak located at $-\nu_i$. The $A_i$ are subtracted from the spectrum $D_i$ and the resulting 
spectrum $D_{i+1}=D_i-A_i$ is the input in the next iteration step.

The factor $g$ in Eq. (\ref{gdef}) is called the {\it gain factor} and takes care of the possibility that the 
highest peak itself could be an artifact. Therefore only a fraction ($g$) of the power is used to calculate the alias
powers. 
The iteration process stops after the $N$th iteration if the power of the highest peak in the subtracted spectrum $D_{N+1}$
is below a given level $P_{min}$ or if a maximum number $N_{max}$ of iteration steps is exceeded. The last output
$D_{N+1}$ is called the {\it residual spectrum}. We used the values $g=0.01$, $P_{min}=0.001$ and $N_{max}=300$.
Normally the peak power in the {\it residual spectrum} falls below this limit of $P_{min}=0.001$ before the maximal
iteration of $N_{max}=300$ is reached.

During the iteration process any artifact should be removed and the main output is the {\it clean component spectrum} 
which is defined as 
\begin{equation}\label{cleancomps}
 C(\nu) = \sum_{i=1}^N \int_{-\infty}^{+\infty} c_i~\delta(\hat{\nu} - \nu_i)~d\hat\nu ,
\end{equation} 
where $\delta (x)$ is the Dirac delta function and the $c_i$ are the clean components from Eq. (\ref{gdef}). The 
resolution of the {\it clean component spectrum} in Eq. (\ref{cleancomps}) is given by the artificial frequency spacing 
$\Delta\nu$ used in the algorithm (see above). Since the resolution of the spectrum $F$ that results from a discrete data
set is $\Delta\nu \simeq 1/T$ we have to smooth the clean component spectrum with a beam function $B(\nu)$. This is 
necessary because the power of the peaks in the spectra $D_I$ is distributed into $n_B=\Delta\nu /T$ frequency bins 
({\it points per beam}) in the interval $[\nu_i-1/T,\nu_i+1/T]$. 
The function $B(\nu)$ is found by a fit of the main peak of the window function at $\nu=0$.

The final step of the CLEAN algorithm is the calculation of the {\it clean spectrum} $S(\nu)$ that contains the 
periodogram of the signal where the window function $W(\nu)$ has been removed. $S(\nu)$ represents the spectrum $F(\nu)$
in Eq. (\ref{convol}). $S(\nu)$ is obtained by adding the {\it residual spectrum} $D_{N+1}(\nu)$ and the smoothed 
{\it clean component spectrum} $C(\nu)$:
\begin{equation}
S(\nu) = D_{N+1}(\nu) + C(\nu).
\end{equation}



\end{document}